\documentclass[12pt]{article}

\usepackage{graphics}
\usepackage{amsmath}
\usepackage{amsfonts}
\usepackage{amssymb}
\usepackage{graphicx}
\usepackage{rotating}

\begin{document}

\begin{titlepage}
\begin{flushright}
WISC-TH-99-353
\end{flushright}
\begin{center}
\bf\large
{GENERALIZED RANDOM PHASE APPROXIMATION: \\
Zero and Non-Zero Temperature Properties of an Interacting Electron Gas}
\end{center}
\vspace{0.2 in}
\begin{center}
{\bf{A. REBEI}\footnote{e-mail:rebei@math.wisc.edu}}  \\
\it{Department of Physics \\
University of Wisconsin-Madison \\
		Madison, WI 53706}
\end{center}
\vspace{0.15 in}
\begin{center}
\bf{W.N.G. HITCHON}\footnote{e-mail:hitchon@cae.wisc.edu} \\
\it{Materials Science Program  \\
University of Wisconsin-Madison \\
		Madison, WI 53706}
\end{center}

\begin{abstract}
Correlated systems at both zero and nonzero temperature
are treated here from a novel angle using  a
functional method. This functional method is an extension of the usual
effective potential method. Here, however
the effective action is made to depend
explicitly on the correlation effects that are inherent in the
physics involved.
This will enable us to obtain  new expressions for the free energy, the
specific heat and the ground state energy.
The new expansion
is shown to give the expected results for the homogeneous case at zero
temperature.
However at non-zero
temperature we are able to get new sets of
diagrams that have a vanishing effect
at zero temperature. To lowest order these
diagrams if summed properly will solve
a $\ln T$ anomaly in the specific heat of an electron gas at
low temperature. We are also able to
show that this method provides a very clear way to extend the RPA
approximation. We  calculate the effect of exchange on the ring diagrams
at zero temperature and
show how to include some of the ladder diagrams. Our results agree
well with known numerical calculations. We conclude by
showing that this method is in fact a variant
of  the time dependent density functional method and
can in principle be applied to study the effects of
correlation in the nonhomogeneous case.
\end{abstract}

\vspace{0.3 in}

\textbf{PACS :}71.10.Ca, 05.30.-d, 05.30.Fk, 31.15.Gy, 71.27.+a, 71.15.Mb

\end{titlepage}

\section{Introduction}

Strongly correlated systems continue to be a popular subject due to the
importance of these systems in physical processes in general. Electron systems
form an important subset of these systems. The nonlinearity that appears in
this type of problem complicates greatly the calculation of any physical
properties of the system that strongly depend on the correlation between the
individual particles. In this paper we are mainly 
concerned with a particular aspect
of this area, that is, of calculating the ground state equilibrium energy of a
large Fermi gas in an external potential, at both zero and at nonzero
temperature. At zero temperature the random phase approximation (RPA) is a
well known method which addresses this problem. Here we present a way to
improve on it, using an effective action method, which includes higher order
interactions in a consistent way. This new method is also applicable at finite
temperature. At low temperature, the \ new method shows clearly the importance
of the so called ``anomalous diagrams'' in cancelling the non-physical $\ln T$
term in the specific heat \cite{isihara}. Another 
important aspect of this new
approach is that it is applicable to nonhomogeneous Fermi systems. The
currently popular method for treating such systems is the density functional
method (DFT)\cite{kohn}, \cite{sham}. In DFT, the 
density of the system
plays a central role and the energy is a functional of it. Both the current
method and density functional method are extensions of the Thomas-Fermi
method. In DFT the density is found by solving self-consistently a
one-particle Schrodinger equation. Correlation effects are taken into account
by using the homogeneous result locally. The kinetic energy term is taken to
be that of a non-interacting system having the same density as the real system
under consideration. In spite of all these approximations, density functional
methods have proved to give better values for binding energies than
Hartree-Fock in atoms, molecules and solids in general. However the method
works best only for almost homogeneous systems; surface effects are handled
poorly by this method. So far the attempt to include a gradient of the density
in the energy functional makes the result worse. The success of the method is
therefore limited. Here we propose a different method, that in principle can
treat nonhomogeneity in a better fashion. However, the problem becomes harder
to manage numerically.

The method proposed here is a path integral method known as the effective
action method. In one dimension it reduces to the well-known WKB method to
first approximation. This work is a first step toward finding a new practical
scheme for the evaluation of correlation energies and other physical
properties of strongly correlated electron systems using this method.

In the homogeneous case, the effectiveness of the method was demonstrated in
Ref.\cite{rebei}. A closely related approach also appears in Ref.
\cite{weinberg} where the gap equation in a superconductor has been derived
using effective action ideas. In
field theory this method also received some attention recently in treating a
scalar field with a quartic interaction \cite{mossB}. Here we 
carry out the
calculations\ of the energy in the non-homogeneous case as far as possible
analytically. It will be seen that the method provides a very compact
expression for the energy that goes beyond the RPA method in a natural way. We
also apply our result for the energy to the homogeneous case. We \ explicitly
calculate the effect of taking exchange into account in the ring diagrams. Our
results agree well with results obtained using quantum Monte Carlo methods.

In the following, we will present a way to compute the free energy of the
electron system. Zero temperature results follow easily from this
calculation. The system is assumed to be in an external static potential
$V(\vec{x})$. The result that we get is very general and could be applied to
many different systems.

\begin{figure}[ptb]
\resizebox{\textwidth}{!}
{\includegraphics[0in,0in][8in,9in]{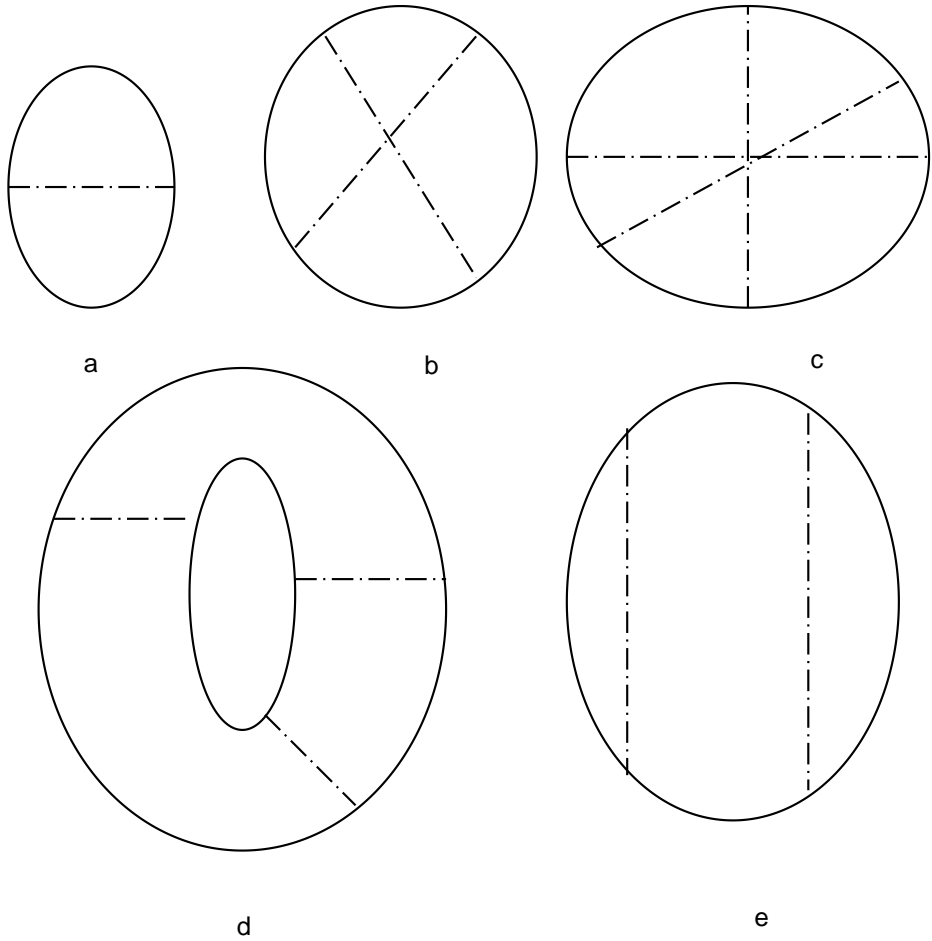}}\caption{Graphs a, b, c and d
are part of the expansion terms in $\Gamma$. Graph e is however reducible and
does not show up in the expansion. The solid line represents the propagator
$\rho(x,y)$, the dashed line represents the propagator $C(x,y)$.}%
\label{expansion}%
\end{figure}

The paper is laid out as follows: In the next section, section 2, we introduce
the thermodynamic potential $\Omega$ for a system subject to external sources.
$\Omega$ therefore will be a functional of these sources. The functional
$\Omega$ is the generator of connected Green's functions at finite
temperature. Usually, one point source functions are used. Here, we will
instead introduce two-point functions \cite{domincis}, \cite{vasilev},
\cite{cornwall},\cite{hawking}. The introduction 
of two-point external sources
enables us to take into account the higher order corrections to the stationary
phase approximation of the partition function in a better way. By including
merely two diagrams (expressed in terms of the variables which are conjugate
to the external sources) in the effective action, we are able to obtain the
contribution of all the second order diagrams (and beyond) of the Coulomb
interaction to the energy in a very compact way (see Fig.~\ref{expansion}).
This is the crucial advantage of this version of the effective action. The
diagrams which are included here are two-particle irreducible, in contrast to
the one-particle irreducible diagrams that we get when only one-point
functions are used in the effective action. However, before doing this, we
introduce a new field to replace the quartic Coulomb interaction by the well
known Stratanovich-Hubbard transformation, so that all the interactions become
local. This new field is simply the Hartree potential as in the RPA
method \cite{negele}. However the treatment presented here is somewhat
different from the one in the aforementioned reference. Here, we explicitly
introduce a term that describes correlation of the Hartree field at two
locations. The sources can also be taken to be instantaneous so we can get a
time-translation invariant solution.

In section 3, we introduce the finite temperature 
effective action $\Gamma$. We get this action by a Legendre 
transformation from $\Omega$. Therefore, we
introduce new variables conjugate to the sources. This is exactly the same
type of transformation as is used in thermodynamics. At the tree level, the
$\Gamma$ functional is simply the action of the system and is related to the
free energy of the system. For homogeneous systems, $\Gamma$ is known as the
effective potential. In our case $\Gamma$ will be a functional of three
variables, the Hartree potential $\varphi(\vec{r})$, the correlation function
$C(x,y)$ of the Hartree potential, and the Green's function $\rho_{\alpha
\beta}(x,y)$ of the Fermi field. If the external sources are taken to be
instantaneous, the points $x = (\tau_{1},\vec{r_{1}}) $ and $y = (\tau
_{2},\vec{r_{2}})$, correspond to the same temperature, $\tau_{1}=\tau_{2}$,
and the function $\rho_{\alpha\beta}(x,y)$ simply becomes the density matrix
of the system. $\alpha$ and $\beta$ are spin indices. Even though it is
possible to express the free energy in terms of the density matrix, it makes
the manipulations more cumbersome. 

In section 4, we solve for $\varphi(x)$ and $C(x,y)$ in terms of $\rho
_{\alpha\beta}(x,y)$. We are able to do this simply because in reality
$\rho_{\alpha\beta}(x,y)$ is the only independent variable. In this case we
get an expression for the effective action $\Gamma$ in terms of the density
only. Therefore getting an expression for $\Gamma$ in terms of $\rho
_{\alpha\beta}(x,y)$ is the essential result of this work from which other
important calculations can be made.

In section 5, we give an explicit expression for the free energy in the
homogeneous case in a neutral background. Here we go beyond the RPA results.
We show that new diagrams appear at finite temperature. The famous `anomalous'
diagram is just one of them  (see Fig.~\ref{anomalous}). An important
outcome of this calculation is the disappearance of the $lnT$ behavior
 in the specific heat that is due to exchange \cite{akira} .

In section 6, we treat the zero temperature case. We show explicitly that the
anomalous diagrams disappear as they should. Here again we go beyond RPA to
include exchange effects on the ring diagrams. We also show that some of the
ladder terms are included in our approximation.

In section 7, we continue the treatment started in the previous section and
calculate the contribution of the ring diagrams if they include exchange.

In the conclusion, we reexamine the non-homogeneous problem. We show how this
method is related to the DFT formalism. In fact it is shown that a statement
like the Hohenberg-Kohn theorem is trivially realized in this formalism.
Similarly the question of v-representability can be given an affirmative
answer within perturbation theory.

\newpage

\section{The Thermodynamic Potential $\mathbf{\Omega}$}

In this section, we obtain an expression for the functional $\Omega$ to
`order' $\hbar^{2}$. This functional is the logarithm of the partition functional
$Z$ of the system in the presence of external sources. The $Z$ functional is
written as an integral over all possible allowable paths of the different
fields with a weighting factor that depends on the value of the action of the
given path. The functional $\Omega$ is simply the generating functional of
connected Green's functions at finite temperature of the fields involved.
However because the Coulomb interaction, quartic in the Fermion field, is hard
to integrate, we introduce an auxiliary field by means of the well known
Hubbard-Stratanovich transformation so that we can avoid the quartic
interaction and its nonlocal behavior. In other words, the procedure consists
of transforming the problem into one in which the electrons interact locally
with a Hartree-type potential. In this treatment, exchange energy terms will
show up directly because of the anti-commutation of the Fermi fields, which is
built into the calculation as will be described below. The correlation terms
are due to the quantum fluctuations around the Hartree potential.

In the following we give an explicit derivation of the terms to order $\hbar^{2}$
in $\Omega$ and no counting will be necessary. What we mean here is that we
are including effects of second order in the \emph{full} Hartree field of the
theory. The equation of motion satisfied by $\rho(x,y)$ goes therefore beyond
the Hartree-Fock approximation and involves second order exchange effects.

\noindent A non-relativistic interacting electron gas in an external potential
$V(\vec{x})$ has the following Hamiltonian in the second quantized form:%

\begin{align}
H_{0}  & =\int d^{3}x\left[  -\frac{1}{2}\Psi_{\alpha}^{\dagger}(x)\nabla
^{2}\Psi_{\alpha}(x)+V(x)\Psi_{\alpha}^{\dagger}(x)\Psi_{\alpha}(x)\right]
\nonumber\\
& +\frac{1}{2}\int d^{3}xd^{3}y\frac{1}{|\vec{x}-\vec{y}|}\Psi_{\alpha
}^{\dagger}(x)\Psi_{\nu}^{\dagger}(y)\Psi_{\nu}(y)\Psi_{\alpha}(x).
\end{align}
\newline Here, we use units such that $\hbar=m=e=k_{B}=1$ and $\beta
\equiv\frac{1}{T}$. $\Psi(x)$ is a two-component temperature-dependent
electron field. $\alpha$ and $\nu$ are spin indices, i.e., $\alpha=1$ for spin
up and $\alpha=-1$ for spin down. Summation is implicit for repeated indices.
In the following $\vec{x}$ will always mean a 3-D space vector. The system is
constrained by the condition%

\begin{equation}
\int d^{3}x\Psi_{\alpha}^{\dagger}(x)\Psi_{\alpha}(x)=N
\end{equation}
where N is the electron number operator which is constant. Therefore the
density operator is simply $\Psi_{\alpha}^{\dagger}(x)\Psi_{\alpha}(x).$
Because of this constraint, we prefer to work instead with the following Hamiltonian%

\begin{equation}
H=H_{0}-\mu N
\end{equation}
where $\mu$ is a Lagrangian multiplier. The electron operator $\Psi(x)$ can be
expanded in terms of a complete orthonormal set of one-particle functions
$\varphi_{k}(x),$ so we write
\begin{align}
\Psi(x)  & =\sum_{k}a_{k}\varphi_{k}(x)\text{ },\nonumber\\
\Psi^{\dagger}(x)  & =\sum_{k}a_{k}^{\dagger}\varphi_{k}^{\ast}(x)\text{ }.
\end{align}
$a_{k}$ and $a_{k}^{\dagger}$ are annihilation and creation operators. The
subscript $k$ includes momentum and spin. The ground state and the excited
states can be represented as Slater determinants formed by the wave functions
$\varphi_{k}(x)$. In our case, it will be advantageous to take these functions
to be self-consistent Hartree eigenfunctions. In the homogeneous case they
reduce to plane waves. Since we are going to use a path integral formulation,
we give the Lagrangian associated with the Hamiltonian H.%

\begin{align}
L  & =  \int\! d^{3}x\left[  i\Psi^{\dagger}(x) \frac{\partial
}{\partial t}\Psi(x)+\frac{1}{2}\Psi(x)\nabla^{2}\Psi^{\dagger}(x)-V(\vec
{x})\Psi^{\dagger}(x)\Psi(x)+\mu\Psi^{\dagger}(x)\Psi(x)\right] \nonumber\\
& -\frac{1}{2}\int d^{3}xd^{3}y\frac{1}{|\vec{x}-\vec{y}|}\Psi^{\dagger
}(x)\Psi^{\dagger}(y)\Psi(y)\Psi(x) .
\end{align}
\newline So for a system at thermal equilibrium, we have to replace the time
component t in the Hamiltonian by $-i\tau$. The partition function $Z$ is as
usual defined as a sum over all possible states $\Phi$. In the absence of
external sources, we have%

\begin{align}
Z[0]  & =\sum_{\Phi} \left\langle {\Phi} \, \vert\, e^{-\beta H}\, \vert\,
{\Phi}\right\rangle \nonumber\\
& =N\int D\Psi D{\Psi}^{\dagger} \exp\left( {-S[\Psi,\Psi^{\dagger}]} \right)
\end{align}
where N is a normalization constant and the integration measure of the Fermi
fields is only defined for fields that satisfy:
\begin{equation}
\Psi(0,x) = - \Psi(\beta,x) .
\end{equation}
$S[\Psi,\Psi^{\dagger}]$ is the action of the system%

\begin{align}
S[\Psi,\Psi^{\dagger}]  & =\int_{0}^{\beta}d\tau\,L\nonumber\\
& =\int_{0}^{\beta}d\tau\int d^{3}x\;\left \{ \Psi^{\dagger}(x)\left [ \frac
{-\partial}{\partial\tau}+\frac{1}{2}\nabla^{2}+\mu-V(x)\right ]
\Psi(x) \right \}
\nonumber\\
& -\frac{1}{2}\int_{0}^{\beta}d\tau\int d^{3}xd^{3}y\;\frac{1}{|\vec{x}%
-\vec{y}|}\Psi^{\dagger}(x)\Psi^{\dagger}(y)\Psi(y)\Psi(x).
\end{align}\\
 The functional integral for the partition function becomes

\begin{eqnarray}
Z[0]  & =& N\int D\Psi D\Psi^{\dagger}\exp \left \{ -\int d\tau_{x}
d^{3}x \left [ \Psi^{\dagger}(x)\left(  \frac{\partial}{\partial\tau}-\frac
{1}{2}\nabla^{2}-\mu+V(\vec{x})\right)  \Psi(x) \right . \right . \nonumber\\
& &\mbox{} \left . \left . + \frac{1}{2}
\int_{0}^{\beta}d\tau\;d\tau_{y}\int d^{3}y
\frac{\Psi^{\dagger}(x)\Psi(x)\Psi^{\dagger}(y)
\Psi(y)}{|\vec{x}-\vec{y}|}\delta(\tau-\tau_{y}) \right ] \right \}.
\end{eqnarray}
In writing the last term we made use of the fact that the Fermi fields satisfy
a Grassmann algebra. An infinite self-energy term has been dropped from the
above expression. Such a term cancels at the end and has no effect on the
evaluation of the eigenfunctions or related physical quantities. Now, we set
for notational convenience:%

\begin{equation}
G^{-1}(x,y)=\left(  \frac{\partial}{\partial\tau_{x}}-\frac{1}{2}\nabla
^{2}-\mu+V(\vec{x})\right)  \delta(x-y)
\end{equation}
and%

\begin{equation}
A(x,y)=\frac{\delta(\tau_{x}-\tau_{y})}{|\,\vec{x}-\vec{y}\,|}\text{ }.
\end{equation}
\newline Before proceeding further, we have to get rid of the quartic term in
the Lagrangian. This is done by introducing an auxiliary boson field.\newline
We write%

\begin{align}
Z[0]  & =N^{\prime}\int D\Psi D\Psi^{\dagger}D\varphi\;\exp \left \{-\int
d^{4}xd^{4}y\;\Psi_{\alpha}^{\dagger}(x)G^{-1}(x,y)
\Psi_{\alpha}(y) \right .\nonumber\\
&\left . -\frac{1}{2}\int d^{4}xd^{4}y\;\varphi(x)A^{-1}(x,y)\varphi(y)+\int
d^{4}x\;\varphi(x)\Psi_{\alpha}^{\dagger}(x)\Psi(x) \right \}.
\end{align}
\newline The fourth component is the time component integrated over the proper
range. It is easy to see that by using the following formula,%

\begin{equation}
\int dx\;e^{-\frac{1}{2}xQx+bx}=(\det Q)^{-\frac{1}{2}}\;e^{\frac{1}{2}%
bQ^{-1}b}%
\end{equation}
\newline and integrating over $\varphi$ we get back the original expression
for $Z\lbrack0 \rbrack$. The prefactor $N^{\prime}$ is a new normalization
constant. The operator $A(x,y)$ is clearly invertible,%

\begin{equation}
A^{-1}(x,y)=-\frac{1}{4\pi}\delta(x-y)\nabla^{2}\text{ }.
\end{equation}
\newline The non-local character of the Coulomb interaction has been removed
by introducing the new bosonic field $\varphi(x).$ It can be shown that
$\varphi(x)$ is the Hartree potential by using the new equations of motion
derived from the new action of the problem. We also couple the fields to local
and non-local sources $J(x),Q(x,y)$ and $B(x,y)$. Now we introduce the
functional $\Omega$, a generator of connected Green functions. This is
introduced through the normalized partition functional which is now a
functional of the external sources. It is given by:%

\begin{equation}
Z[J,B,Q] =\exp\left \{ -\beta\;\Omega[J,B,Q] \right \}
\end{equation}
such that

\begin{eqnarray}
\lefteqn{\exp\left(-\Omega\lbrack J,B,Q \rbrack\right)\int D\Psi D\Psi^{\dagger}D\varphi
\;\exp\left(  -S[\Psi,\Psi^{\dagger},\varphi]\right)   = } \nonumber\\
& & \int D\Psi D\Psi^{\dagger}D\varphi\exp \left \{ -S[\Psi,\Psi^{\dagger}
,\varphi]+\int d^{4}xd^{4}y\Psi^{\dagger}(x)Q(x,y)\Psi(y) \right . \nonumber \\
& &\mbox{} + \left . \frac{1}{2}\int d^{4}xd^{4}y\varphi(x)B(x,y)\varphi(y) 
+\int d^{4}x\varphi(x)J(x) \right \}
\end{eqnarray}
\newline where we have set $\beta = 1$.  
$S[\Psi,\Psi^{\dagger},\varphi]$ is simply the action of the
transformed problem,%

\begin{align}
S[\Psi,\Psi^{\dagger},\varphi]  & =\int d^{4}xd^{4}y\;\Psi^{\dagger}%
(x)G^{-1}(x,y)\Psi(y)+\frac{1}{2}\varphi(x)A^{-1}(x,y)\varphi(y)\nonumber\\
& -\int d^{4}x\varphi(x)\Psi^{\dagger}(x)\Psi(x)
\end{align}

Now we define three new variables:%

\begin{equation}
\frac{\delta\Omega[J,B,Q]}{\delta J(x)}\vert_{J=B=Q=0}=\left\langle
\varphi(x)\right\rangle \equiv\varphi_{c}(x)
\end{equation}%

\begin{equation}
\frac{\delta\Omega\lbrack J,B,Q]}{\delta B(x,y)}|_{J=B=Q=0} =\frac{1}
{2}\left\langle \varphi(x)\varphi(y) \right\rangle \equiv\frac{1}{2}
\left [ \varphi_{c}(x)\varphi_{c}(y)+C(x,y) \right ]
\end{equation}%

\begin{equation}
\frac{\delta\Omega\lbrack J,B,Q]}{\delta Q(x,y)}|_{J=B=Q=0}=\left\langle
\Psi_{\alpha}^{\dagger}(x)\Psi_{\beta}(y)\right\rangle \equiv\rho_{\alpha
\beta}(x,y)
\end{equation}
\newline $\varphi_{c}(x)$ is the expectation value of the field $\varphi(x)$
in the ground state. $C(x,y)$ is a correlation function of the field
$\varphi(x).$ $\rho_{\alpha\beta}(x,y)$ is the Green function of the Fermi
field. $\rho_{\alpha\beta}(x,x)$ is therefore the density of the system. Here
and below, the ``time'' ordering operator is not written explicitly. Therefore
terms like $\rho(\vec{x},\vec{y})$ are defined by setting $\tau_{x}-\tau
_{y}=0^{+}.$ Note that $C(x,y)$ measures the departure from quasi-independence
due to the correlation between the values of the potential at two different
locations. The expectation value of the Fermi field $\Psi(x)$ is zero. In the
following, we will obtain an explicit expression for $\Omega\lbrack J,B,Q]$ to
`order' $\hbar^{2}$. Then 
by solving for $\varphi_{c}(x),C(x,y)$ and $\rho(x,y)$
in terms of $J(x),B(x,y)$ and $Q(x,y)$ we get an expression for the effective
action $\Gamma$ which is a functional of the new variables. This will be done
in the next section.\newline Now we turn to determining $\Omega$. First we
expand the exponent in the above integral around $\Psi=\Psi^{\dagger}=0$ and
$\varphi=\varphi_{0}$ where $\varphi_{0}$ is the configuration of $\varphi$
that extremizes the action S. Therefore, we have%

\begin{equation}
\left(  A^{-1}+B\right)  \varphi_{0}=-J .
\end{equation}
\newline It is understood from the above that there is an integration over
space and time on the L.H.S. of this expression. We choose from now on not to
write integrals explicitly unless there might be some confusion. Now we expand
around $\varphi_{0}$, so we write%

\begin{equation}
\varphi_{(old)}=\varphi_{(new)}+\varphi_{0} .
\end{equation}
\newline Then assuming that the main contribution to the integral comes from
the saddle point, we get the following 
expression for the partition functional:

\begin{eqnarray}
\exp\{-\beta\Omega\}&&=\exp(-\beta\;S[\varphi_{0}])\;\frac{\det[\widetilde
{G}^{-1}+Q]\;\det[A^{-1}+B]^{-\frac{1}{2}}}{\det[G^{-1}]\;\det[A^{-1}
]^{-\frac{1}{2}}}\,\Sigma^{-1} \nonumber  \\
&& \mbox{} \times\,\exp \left [ \int D\Psi D\Psi^{\dagger}D\varphi\;\;\Xi\; 
\left \{\frac{1}{2}e^{2}(\varphi\Psi^{\dagger}\Psi)^{2}\,\;+\ldots \right \}
\, \right ]
\end{eqnarray}
where we have defined $\Xi$ and $\Sigma$ to be
\begin{align}
\Xi & =\exp \left [-\Psi^{\dagger}(\widetilde{G}^{-1}+Q)\Psi-\frac{1}%
{2}\varphi(A^{-1}+B)\varphi \right ] \\
\Sigma & =\int D\Psi D\Psi^{\dagger}D\varphi\;\exp \left [-\Psi^{\dagger
}(G^{-1}+Q)\Psi-\frac{1}{2}\varphi(A^{-1}+B)\varphi \right ]
\end{align}
\newline In the above, we have used the fact that%

\begin{equation}
\int da\text{ }da^{\dagger}\;e^{-a^{\dagger}Ma}=det\;M
\end{equation}
for Grassmann numbers a and a$^{\dagger}$, and%

\begin{equation}
\int dx\; e^{-\frac{1}{2}xMx}={\left(  det M \right) }^{-\frac{1}{2}}%
\end{equation}
for c-numbers x. The argument of the first term on the right is%

\begin{equation}
S[\varphi_{0}]= \frac{1}{2}\varphi_{0}A^{-1}\varphi_{0} + \frac{1}{2}
\varphi_{0}B\varphi_{0} + J\varphi_{0}%
\end{equation}
and%

\begin{equation}
\widetilde{G}^{-1}=G^{-1}-\varphi_{0} .
\end{equation}
\smallskip Now using the fact that%

\begin{equation}
\det A=e^{Tr\ln A},
\end{equation}
we can get an explicit expression for $\Omega[J,B,Q]$:%

\begin{align}
\Omega\lbrack J,B,Q]  & =S(\varphi_{0})-\;Tr\ln[1+G\,\varphi_{0}+GQ]+\frac
{1}{2}\;Tr\ln[1+A\,B]\nonumber\\
& -\frac{1}{2}\!\int\!D\Psi D\Psi^{\dagger}D\varphi\;(\varphi\Psi^{\dagger
}\Psi)^{2}\;\frac{\Xi}{\Sigma}+\ldots
\end{align}
\newline After finding $\Omega$, we now calculate the effective action
$\Gamma$. We have to solve for the external sources in terms of the physical
variables $\rho$, $C$ and $\varphi_{c}$. This we do in the next section where
we find the effective action at finite temperature. However, this treatment
applies to zero temperatures too. In this case too, we get 
an expansion in $\hbar$
. Again $\hbar$ is not a true expansion parameter in the true sense of the
word. It is only used as a bookkeeping method for the diagrams included in the
expansion of $\Omega$.

\newpage

%

\section{ The Effective Action $\mathbf{\Gamma}$ }

The effective action is a functional of $\varphi_{c}(x)$, $C(x,y)$, and
$\rho(x,y)$, which is obtained by a triple Legendre transformation from
$\Omega\lbrack J,B,Q]$%

\begin{align}
\Gamma\lbrack\varphi_{c},C,\rho]  & =\Omega\lbrack J,B,Q]-\int d^{4}%
x\;\varphi_{c}(x)J(x)-\frac{1}{2}\int d\text{ }[xy]\;\varphi_{c}%
(x)B(x,y)\varphi_{c}(y)\nonumber\\
& -\frac{1}{2}\int d[xy]\;C(x,y)B(x,y)-\int d\text{ }[xy]\rho(x,y)Q(x,y)\;.
\end{align}
where we have written $d$ $[xyz...]$ for $d^{4}xd^{4}yd^{4}z...$ to simplify
the notation. It is understood that the integration over time is carried for
$\tau$ in $[0,\hbar\beta]$ for the nonzero temperature case and over all time
for the zero temperature case. In this section we will assume we are in the
zero temperature regime. However this discussion applies equally well to the
nonzero temperature case.

It is easily verified that :%

\begin{equation}
\frac{\delta\Gamma[\varphi_{c},C,\rho]}{\delta\varphi_{c}(x)} =-J(x)-\int
d^{4}y\; B(x,y)\varphi_{c}(y)
\end{equation}%

\begin{equation}
\frac{\delta\Gamma\lbrack\varphi_{c},C,\rho]}{\delta C(x,y)}=-\frac{1}%
{2}B(x,y),
\end{equation}
and
\begin{equation}
\frac{\delta\Gamma\lbrack\varphi_{c},C,\rho]}{\delta\rho(x,y)}=-Q(x,y)\text{
}.
\end{equation}
\newline When we turn off the external sources, the above equations give the
values of $\varphi_{c},C$ and $\rho$ that minimize the effective action.
$\varphi_{c}(x)$ and $C(x,y)$ are really dependent variables since they depend
on the auxiliary field $\varphi(x)$, so we should be able in principle to
express them in terms of $\rho(x,y)$ which gives us the density.

Now we have to express $J,B$ and $Q$ in terms of $\varphi_{c}$, $C$ and $\rho
$. First we note that when $\hbar=0,$ we have
\begin{equation}
\frac{\delta\Omega}{\delta J}=\varphi_{0} .
\end{equation}

Hence, we can write%

\begin{equation}
\varphi_{0}=\varphi_{c}+\widetilde{\varphi}%
\end{equation}
\medskip where $\widetilde{\varphi}$ is of order $\hbar$. Similarly, we write

\begin{equation}
S(\varphi_{0})=S(\varphi_{c})+ \hbar S_{1} .
\end{equation}
\medskip Now we seek an expression for $\Gamma$ in the form%

\begin{equation}
\Gamma=\Gamma_{0}+ \hbar \Gamma_{1} +\hbar^{2} \Gamma_{2} .
\end{equation}
\medskip It then follows that%

\begin{equation}
\Gamma_{0}=\frac{1}{2}\varphi_{c}A^{-1}\varphi_{c}.
\end{equation}
\medskip Now, we find approximate expressions for $B$ and $Q$ in terms of
$\varphi_{c},C$ and $\rho$. Using Eq.(31) and Eq.(35), we get 
the following relation:

\begin{equation}
\rho^{-1}G=1+GQ-eG\varphi_{c}+O(\hbar)\text{ }.
\end{equation}
\medskip The expression for $B$ is more involved. Again using Eq.(35) and
treating $\varphi_{0}$ as a functional of $J$ and $B$, we get:%

\begin{equation}
B\,C\,=\,(1-2e\,\rho\,\varphi_{c})\left(  1-2\frac{\varphi_{c}\widetilde
{\varphi}}{C}\right)  ^{-1}\,-A^{-1}\,C\text{ }.
\end{equation}
\medskip Inserting back all these expressions in $\Gamma$, we get for
$\Gamma_{1}$

\begin{equation}
\Gamma_{1}=-Tr\;\ln\;(\rho^{-1}G)+\frac{1}{2}Tr\;\ln(C^{-1}A)+\frac{1}%
{2}Tr\;(A^{-1}C)-Tr\;(\rho\widetilde{G}^{-1})\text{ }.
\end{equation}
\newline The terms of order $\hbar^{2}$ are more involved to treat but again
straightforward. The steps are similar to those in the case of the effective
action with one-point sources only  \cite{jackiw}. Actually, 
it can be shown
that the next terms in $\Gamma_{2}$ are the sum of  two-particle irreducible
diagrams of the theory \cite{vasilev},\cite{cornwall}. Here, we 
sum only the
first two diagrams in this series expansion. Fig.~\ref{expansion} shows the
first four diagrams in this expansion. The fifth diagram is not part of the
asymptotic expansion of $\Gamma$. These first two diagrams enable us to
include first and second order exchange effects. Diagram e is not part of
$\Gamma_{2}$ since it is two-particle reducible. However it is one-particle
irreducible and it is part of the usual effective action \cite{negele},

\begin{align}
\Gamma_{2}  & =-\frac{1}{2}e^{2}\int d[xy]\;\rho(x,y)\rho
(y,x)C(x,y)\nonumber\\
& -\frac{1}{4}e^{4}\int d[xyuv]\;\rho(x,y)\rho(y,u)\rho(u,v)\rho
(v,x)C(x,u)C(y,v).
\end{align}

\begin{figure}[ptb]
\resizebox{\textwidth}{!}
{\includegraphics[0in,0in][8in,8in]{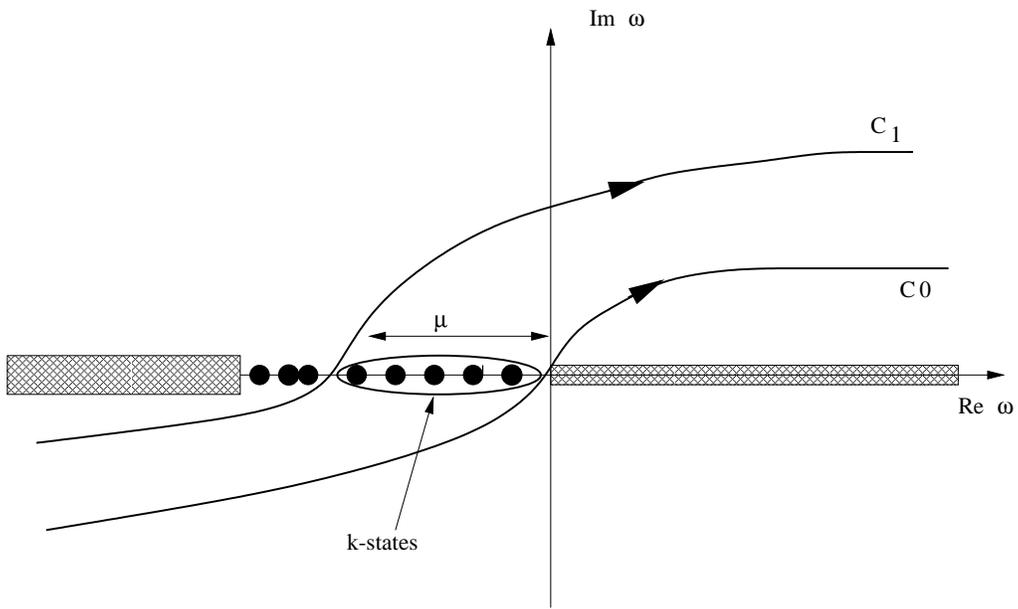}}
\caption{Path of integration used to obtain Eq.(47).}
\label{path}
\end{figure}

\noindent Hence the reducible graphs do not appear in the expansion. It is the
term $\Sigma$ that appeared in Eq.(23) that is responsible for excluding such
a graph from the expansion. $\Gamma_{2}$ and higher order terms represent
vacuum graphs where the propagators are $\rho(x,y)$ and $C(x,y)$ of the Fermi
field and the boson field respectively. Therefore to order $\hbar^{2}$, the
effective action has the following expression%

\begin{align}
\Gamma\lbrack\varphi_{c},C,\rho]  & =-S(\varphi_{c})+\frac{1}{2}Tr\;\ln
AC^{-1}+A^{-1}C\\
& -Tr\;\ln\rho^{-1}G+\widetilde{G}^{-1}(\varphi_{c})\rho+\Gamma_{2}%
[C,\rho],\nonumber
\end{align}
\newline where

\begin{equation}
\widetilde{G}^{-1}(\varphi_{c})=\left(  \frac{\partial}{\partial\tau}-\frac
{1}{2}\nabla^{2}-\mu+V(x)-\varphi_{c}(x)\right)  \text{ }.
\end{equation}
\newline The term $\mu\int d^{3}x$ $\rho(x,x)$ cancels the term $\mu N$ in the
Hamiltonian H, and therefore it can be handled without difficulty. However, we
have to deal with the $\mu$ that appears in the term $Tr\,\ln\rho^{-1}\,G$.
This could be removed by a convenient choice of the path of integration in the
complex $\omega$-plane. The case $\mu=0$ occurs when there are no charges
present. We are dealing with negatively charged electrons bound by a static
potential $V(x)$. Therefore bound states appear, with negative energies
bounded from below for any physically sensible $V(x)$. In Fig.~\ref{path} we
show how by going from the $C_{1}$ path, which corresponds to nonzero $\mu$,
to the $C_{0}$ path we pick up contributions from the bound states between the
two paths, since energies of bound states are poles of the propagator $G^{-1}$
in the complex $\omega$ plane. Note that the eigenvalues $\omega_{k}$ of these
states do not take into account correlations due the Coulomb field. Because of
the logarithmic operator, there is a cut along the real positive axis. By
going back to real time and then Fourier transforming the 
time dependence, we get
\begin{align}
& \int_{C_{1}}dt\;Tr\ln\left(  -i\frac{\partial}{\partial t}-\frac{1}{2}%
\nabla^{2}-\mu+V(x)\right) \nonumber\\
& =\int dt\int_{C_{0}}\frac{d\omega}{2\pi i}\;tr\;\ln\left(  -\omega-\frac
{1}{2}\nabla^{2}+V\right)  \text{ }+\sum_{k}\omega_{k}%
\end{align}
\newline where $tr$ applies only to spatial variables. The $\omega_{k}$'s are
the eigenvalues of the following equation

\begin{equation}
\left(  -\frac{1}{2}\nabla^{2}+V(x)\right)  \varphi_{k}(x)=\omega_{k}%
\varphi_{k}(x)\text{ }.
\end{equation}
\newline The single particle functions $\varphi_{k}(x)$ of the potential
$V(x)$ are not supposed to be used alone as the starting point for any
numerical calculations since they do not take account of the repulsion between
electrons. Any numerical calculations will have to start by solving Eq.(56).
However, it is clearly advantageous to separate the effect of the external
potential $V(x)$. This separation also appears in the usual treatments.
The ground state energy E$_{g}$ is given in terms of the effective action per
unit time. Using the above results, and the fact that the time independent
values of $\rho(x,y),C(x,y)$ and $\varphi_{c}(x)$ are defined by setting
$\tau_{x}=\tau_{y}=0$, the expression for $E_{g}$ is%

\begin{align}
E_{g}\int dt  & =\frac{1}{2}\int d[xy]\;\varphi_{c}(x)A^{-1}(x,y)\varphi
_{c}(y)\;+\frac{1}{2}Tr\;\ln C\;A^{-1}\nonumber\\
& -\frac{1}{2}\int d[xy]\;A^{-1}(x,y)C(y,x)\;-\sum_{\mu<\omega_{k}<0}%
\omega_{k}\int dt\nonumber\\
& -Tr\;\ln\rho G^{-1}+\int d^{4}x\left(  \frac{\partial}{\partial\tau_{x}%
}-\frac{1}{2}\nabla^{2}+V(x)-e\;\varphi_{c}(x)\right)  \rho(x,y)|_{\tau
_{x}=\tau_{y}}^{x=y}\nonumber\\
& +\frac{1}{4}e^{4}\int d[xyuv]\;\rho(x,y)\rho(y,u)\rho(u,v)\rho
(v,x)C(x,u)C(y,v)\nonumber\\
& +\frac{1}{2}e^{2}\int d[xy]\;C(x,y)\rho(x,y)\rho(y,x)
\end{align}
where we have dropped terms of order higher than $e^{4}$ in $\Gamma_{2}$ for
simplicity. This is the full expression for the ground state energy of the
system. Here $G^{-1}$ has $\mu$ set to zero. To be able to use this expression
for $E_{g}$, more approximations are required. In later sections, we use this
expression to find the energy of a homogeneous electron gas both at zero and
nonzero temperature. Since $\varphi_{c}$ is a dependent field, then in
principle we should be able to solve for $\varphi_{c}$ and $C$ in terms of
$\rho$ only. This we do next but only approximately to keep the calculations manageable.

\newpage


\section{The Effective Action as a Functional of $\mathbf{\rho(x,y)}$}

In this section, we find an expression for $\Gamma$ solely in terms of the
Green's function $\rho(x,y).$ This is done by finding the expressions for
$\varphi_{c}(x)$ and $C(x,y)$, that minimize $\Gamma$, in terms of the density
$\rho(x,y).$ For $\varphi_{c}(x)$ we get the following expression:%

\begin{equation}
\varphi_{c}(x)=-e\,\int d^{3}y\,\frac{\rho(y,y)}{|\vec{x}-\vec{y}|}\text{ }.
\end{equation}

\begin{figure}[ptb]
\includegraphics{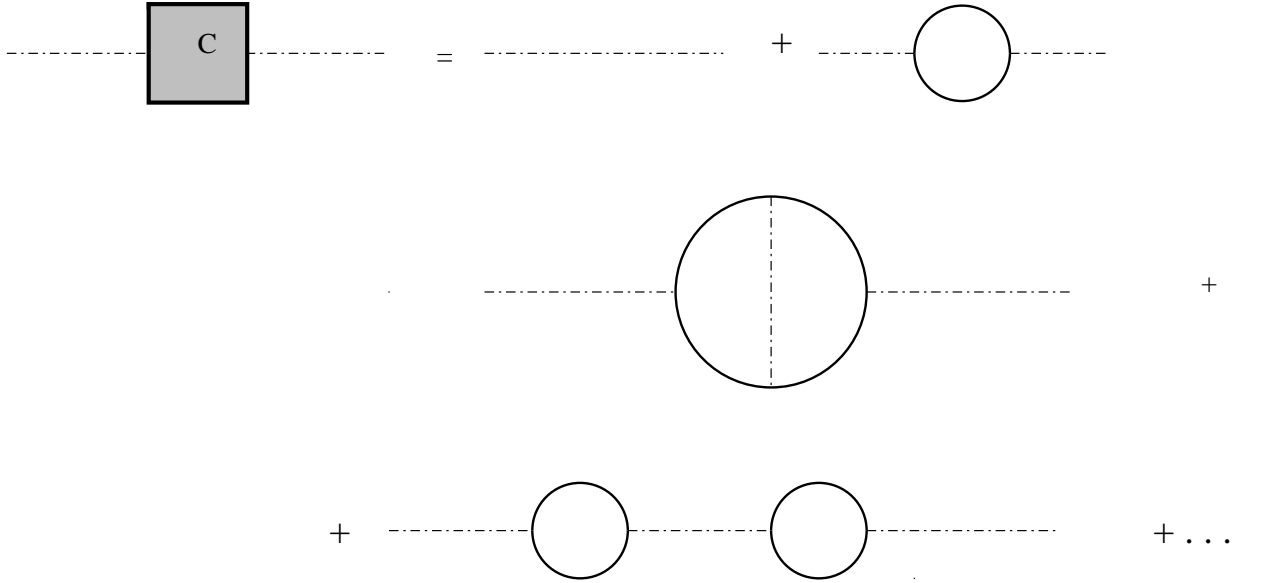}\caption{The expansion of the propagator
$C(x,z)$ to order $e^{4}$ .}%
\label{Cex}%
\end{figure}

\begin{figure}[ptb]
\resizebox{\textwidth}{!}
{\includegraphics[0in,0in][8in,5in]{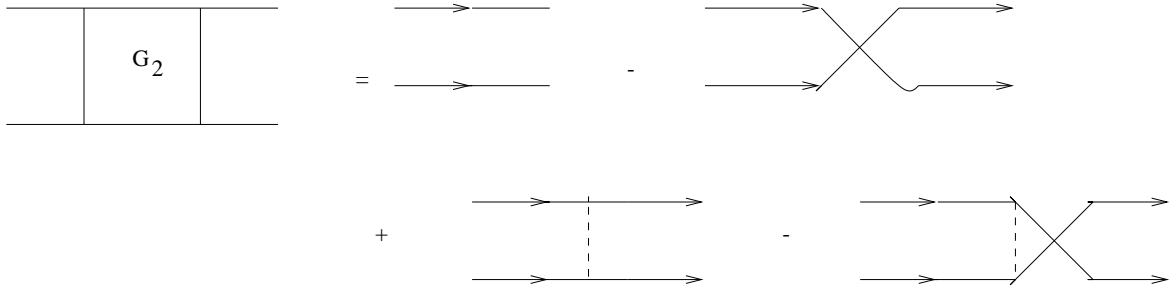}}\caption{Born approximation for
two-particle Green's function. }%
\label{ladder}%
\end{figure}

\begin{figure}[ptb]
\resizebox{\textwidth}{!}
{\includegraphics[0in,0in][8in,5in]{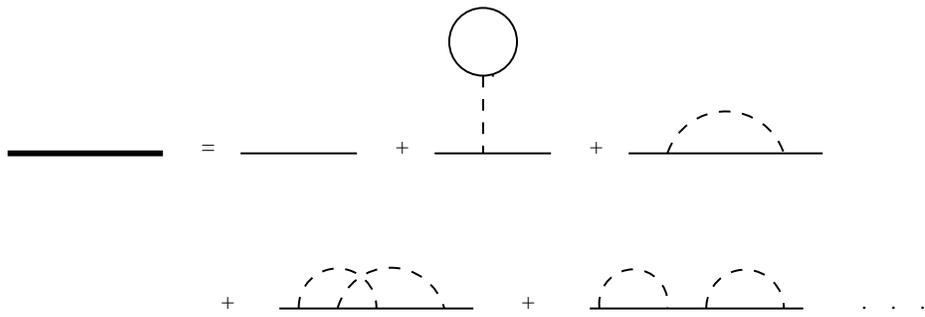}}\caption{Approximate
solution to the one-particle Green's function. The propagators on the 
right correspond to the free theory.}%
\label{rho-approx}%
\end{figure}

\noindent Similarly, minimizing $\Gamma$ with respect to $\rho(x,y)$, we get%

\begin{equation}
\delta(x,z)=\left[  \frac{\partial}{\partial\tau_{x}}-\frac{1}{2}\nabla
^{2}+V(x)-e\;\varphi_{c}(x)\right]  \rho(x,z)+e^{2}\int d^{4}y\,\rho
(x,y)C(x,y)\rho(y,z)
\end{equation}
\begin{equation}
-e^{4}\int d[yuv]\rho(v,u)\rho(u,x)\rho(y,v)\rho(z,y)C(u,y)C(v,x)\: .
\end{equation}
\newline Finally, minimizing $\Gamma$ with respect to C, we get%

\[
\delta(x,z)=\int d^{4}y \left [ A^{-1}(x,y)-e^{2}\rho(x,y)\rho(y,x)\right ]
C(y,z)
\]
\begin{equation}
-e^{4}\int d[yvu]C(u,v)C(y,z)\rho(x,u)\rho(u,y)\rho(y,v)\rho(v,x)\: .
\end{equation}
To be able to solve for $C(x,y)$ we have to linearize the theory. In the
following, we keep terms only to `order' $e^{4}$. Hence, we find that the
correlation $C(x,y)$ is given by%

\[
C(x,z)=A(x,z)+e^{2}\int d[x_{1}x_{2}]\;A(x,x_{1})D(x_{1},x_{2})A(x_{2},z)
\]
\[
+e^{4}\int d[x_{1}x_{2}x_{1}^{\prime}x_{2}^{\prime}]\;A(x,x_{1}^{\prime
})D(x_{1}^{\prime},x_{2}^{\prime})A(x_{2}^{\prime},x_{1})D(x_{1},x_{2}%
)A(x_{2},z)
\]
\begin{equation}
+e^{4}\int d[x_{1}x_{2}x_{1}^{\prime}x_{2}^{\prime}]\;\rho(x_{1},x_{1}%
^{\prime})\rho(x_{1}^{\prime},x_{2})\rho(x_{2},x_{2}^{\prime})\rho
(x_{2}^{\prime},x_{1})A(x_{2},x_{1})A(x_{2}^{\prime},z)A(x,x_{1}^{\prime
})
\end{equation}\\
 where we have set%

\begin{equation}
D(x_{1},x_{2})=\rho(x_{1},x_{2})\rho(x_{2},x_{1})\: .
\end{equation}
\newline The above equation for $C(x,z)$ can be represented diagrammatically
as shown in Fig.~\ref{Cex}. The first term is the bare Coulomb potential.
The second and the third are the direct and exchange term with one bubble. So
far the Fermion propagators are the true propagators in this expansion. In the
following we will ignore the corrections with two bubbles and higher.
Clearly $C(x,y)$ is the full screened Coulomb potential if all
orders of $e^{2}$ are included. Similarly, to order $e^{4}$, the equation
satisfied by $\rho(x,z)$ is%

\[
\left [\frac{\partial}{\partial\tau_{x}}-\frac{1}{2}\nabla^{2}
+V(\vec{x})+e^{2}\int dy\frac{\rho(y,y)}{|\vec{x}-\vec{y}|} \right ]
\rho(x,z)=\delta(x,z)
\]
\[
-e^{2}\int dy\,A(x,y)\rho(x,y)\rho(y,z)-e^{4}\int d[yx_{1}x_{2}]\;\rho
(x,y)\rho(y,z)A(x,x_{1})D(x_{1},x_{2})A(x_{2},y)
\]
\begin{equation}
+e^{4}\int d[yx_{1}x_{2}]\;\rho(x_{2},x_{1})\rho(x_{1},y)\rho(x,x_{2}%
)\rho(y,z)A(x_{1},x)A(x_{2},y)\: .
\end{equation}
This is obviously the known equation satisfied by the Fermi Green's functions
\cite{negele}. The term on the R.H.S. 
of order $e^{2}$ is an exchange term or
Fock term due to the statistics of the electrons. The last two terms of order
$e^{4}$ take into account collisions and are equivalent to the usual Born
approximation for two-particle Green's function in 
scattering theory, Fig.~\ref{ladder}.
 The most likely way to solve these
integro-differential equations is by iteration. Using the above equations, we
can write an explicit expression for the energy in terms of the full
propagator $\rho(x,z)$ of the theory. However, such an expression suffers from
the problem of over-counting some of the states of the system. This is mainly
due to the fact that $e^{2}$ is not a good expansion parameter. Hence physical
arguments are used to discard some terms at this level of approximation and
instead include them at higher orders. An expression for the electron
propagator in terms of the free propagator is given in Fig.~\ref{rho-approx}
. 
The corresponding analytical 
expression can be easily written following
usual rules \cite{negele}. Here we keep 
only the first three terms. The second
term vanishes in the homogeneous case, the case we are interested in this
paper. We expect the expression for the energy, which has not been linearized,
to be a good one if we believe that a stationary phase approximation is
viable. Without including $\Gamma_{2}$, the expression obtained for the energy
is correct with $\rho(x,y)$ the Hartree propagator and $\varphi_{c}(x)$ the
Hartree potential. This indicates that a stationary phase approximation is
possible and that the higher order correction $\Gamma_{2}$ should be a small
perturbation to the Hartree solution. The linearization of the problem must
take account of this. Hence, we can justify the validity 
of the expansion of  $\Gamma$ in $\hbar$,

\begin{equation}
\Gamma=\Gamma_{0}+\hbar \Gamma_{1}+\hbar^{2}\Gamma_{2}+...
\end{equation}\\
by the result we get in the end.
 To second order in $\hbar$, Eq.(54) becomes
\[
\int dyA^{-1}(x,y)\;C(y,z)=\delta(x,z)+\hbar e^{2}\int\: dyD(x,y)A(y,z)
\]
\[
+\hbar e^{4}\int\text{ }d[yuv]\: \rho(x,u)\rho(u,y)\rho(y,v)\rho
(v,x)A(u,v)A(y,z)
\]
\begin{equation}
-2\hbar^{2}e^{4}\int d[yuv]\rho(x,u)\rho(u,v)\rho(v,y)\rho(y,x)A(u,v)A(y,z)
\end{equation}
where $\rho(x,y)$ is the free electron propagator. We have used only a first
order approximation to the true propagator to find this equation for $C(x,y)$.
Using this equation, we get an expression for the energy to order $\hbar^{2}$ with
proper symmetry factors for the diagrams involved in the expansion. Setting
$\hbar=1$, we obtain for an interval of time T,%

\[
T\;E_{g}[\rho]=\frac{e^{2}}{2}\int dx\,dy\;\frac{\rho_{H}(x,x)\rho_{H}
(y,y)}{|\vec{x}-\vec{y}|}\,+\,e\int dx\,V(x)\rho_{H}(x,x)
\]
\[
+\int dx\left(  -\frac{1}{2}\nabla^{2}\rho_{H}(x,y)\right)  |_{x=y}
\;-T\sum_{\omega_{k}>\mu}\omega_{k}
\]
\[
-\frac{1}{4}\;e^{4}\!\int dx_{1}\,dx_{2}dx\,dy\;\rho_{H}(x_{1},x)\rho
_{H}(x,x_{2})\rho_{H}(x_{2},y)\rho_{H}(y,x_{1})A(x,y)A(x_{2},x_{1})
\]
\[
+\frac{1}{2}\;Tr\,\ln \left [\text{ }\delta(x,z)+e^{2}\int dyA(z,y)D_{H}
(x,y)+ \right .
\]
\[
\;e^{4}\int dydx_{1}dx_{2}\;\rho_{H}(x_{1},x)\rho_{H}(x,x_{2})\rho_{H}
(x_{2},y)\rho_{H}(y,x_{1})A(x_{2},x_{1})A(y,z)\;-
\]
\[
\left . 2e^{4}\int dydx_{1}dx_{2}\;\rho_{H}(x,x_{1})\rho_{H}(x_{1},x_{2})\rho
_{H}(x_{2},y)\rho_{H}(y,x)A(x_{2},x_{1})A(y,z)\;\right ]
\]
\[
-Tr\,\ln \left [\;\delta(x,z)-e^{2}\!\int dy\;\rho_{H}(y,y)\rho
_{H}(x,z)A(x,y)-e^{2}\!\int dy\;A(x,y)\rho_{H}(x,y)\rho_{H}(y,z) \right .
\]
\[
-e^{4}\int dydx_{1}dx_{2}\;\rho_{H}(x,y)\rho_{H}(y,z)\rho_{H}(x_{1},x_{2}%
)\rho_{H}(x_{2},x_{1})A(x_{1},x)A(x_{2},y)
\]
\[
\left [+e^{4}\int dydx_{1}dx_{2}\;\rho_{H}(x_{2},x_{1})\rho_{H}(x_{1},y)\rho
_{H}(x,x_{2})\rho_{H}(y,z)A(x_{1},x)A(x_{2},y)\; \right ]
\]
\[
-e^{2}\!\int dy\;\rho_{H}(y,y)\rho_{H}(x,z)A(y,x)|_{x=z}+e^{2}\int
dxdy\;A(x,y)\rho_{H}(x,y)\rho_{H}(y,z)\;|_{x=z}
\]
\[
-e^{4}\!\int dxdydx_{1}dx_{2}\delta(x-z)\;A(x,x_{1})A(y,x_{2})\rho
_{H}(y,z)\rho_{H}(x_{2},x_{1})\rho_{H}(x_{1},y)\rho_{H}(x,x_{2})
\]
\begin{equation}
+e^{4}\int dxdydx_{1}dx_{2}\;\delta(x-z)\rho_{H}(x,x_{1})\rho_{H}(x_{1}%
,x_{2})\rho_{H}(x_{2},y)\rho_{H}(y,x)A(x_{2},x_{1})A(y,z) .
\end{equation}
\newline The trace, $Tr$, acts on x and z. So now we have obtained an
expression for the energy in terms of the Green's functions $\rho(x,z)$ only.
This is a major result of the work. As we mentioned above, we solve for
$\rho(x,y)$ to order $e^{4}$ by iteration of Eqs.(52-53). The expression for
the energy can be solely written in terms of $\rho(x,x)$. However as we stated
earlier this is not advantageous and it is better to keep working in terms of
$\rho(x,y)$.

We see that the classical Coulomb term and the interaction with the external
potential appear naturally in this approximation. They can be easily
separated from the full expression for the energy.
Another important point is
that a gradient of the density also appears naturally within the above
expressions. 
Therefore the kinetic term is treated more accurately in this
method compared to the density functional method. By expanding the first
logarithm, we immediately obtain the Hartree-Fock exchange term, i.e.,%

\begin{equation}
E_{exch}=\frac{1}{2}e^{2}\int dx\,dy\frac{\rho_{H}(x,y)\rho_{H}(y,x)}{|\vec
{x}-\vec{y}|}\delta(\tau_{x}-\tau_{y}).
\end{equation}\\
 The usual expression for the exchange energy follows trivially from
the previous expression upon using the commutation relation of the Fermi
field. The infinity arising from this cancels the self-interaction term that
was dropped in Eq.(9). There is a similar exchange term that comes from the
second logarithmic term . This term is canceled by another similar term
outside the logarithm. The usual RPA term is contained in the first logarithm.
The new, extra term in the same logarithm provides among other things exchange
corrections to the ring diagrams. The higher order exchange diagrams are also
included in this approximation. We will say more on this when we treat the
homogeneous case in the following sections.\newline Before we go on to the
next section, we should point out that an expression for the energy in terms
of the density only is possible in this formalism  \cite{rebeiTh}. The
expression is complicated however and does not seem to be useful. We were also
able to show that a derivative expansion in terms of the Hartree field is more
suitable in our case rather than seeking one in terms of the density. In fact
a derivative expansion in terms of the Hartree field seems more appropriate
since it is better behaved than the density for problems where the density
changes sharply. We hope to address these questions in a later communication.

\newpage

%

\section{The Fermi Gas at Low Temperature}

\begin{figure}[ptb]
\resizebox{\textwidth}{!}
{\includegraphics[0in,0in][8in,8in]{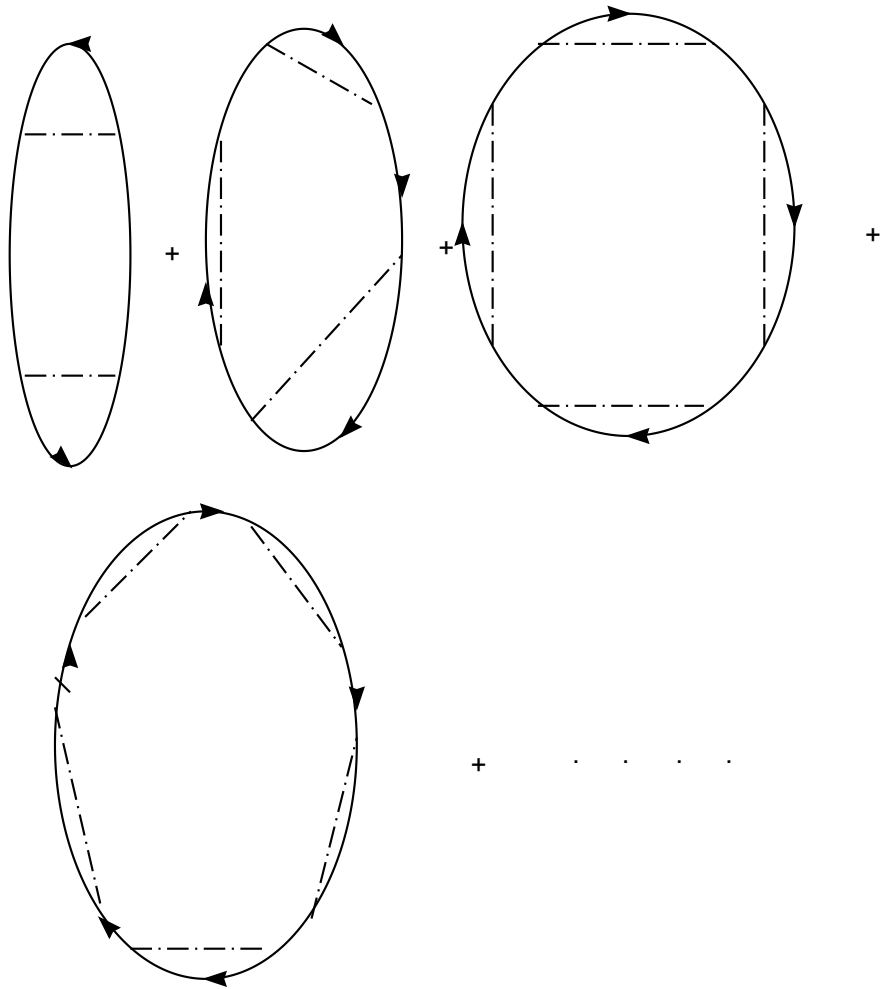}}\caption{A new infinite set of
diagrams at nonzero temperature.}
\label{anomalous}
\end{figure}

In this section we use the result found previously to improve on calculations
of the exchange energy and ring diagrams made by Isihara and coworkers
 and others and calculate the specific heat at 
low temperature \cite{akira} . A
short review of these calculations appears in Ref.\cite{kraeft} 
and offers a good
comparison of all these methods up to the year 1975. The authors of this paper
end with an interesting conjecture which we prove here to be true. The
appearance of temperature dependent logarithmic terms in the internal energy
found through various calculations is unacceptable. These terms will reappear
again in the specific heat and will spoil the observed linear behavior at low
temperature. In the past these difficulties were not given close attention.
They were either hidden in an effective mass or the expansion was not
completely correct as we argue below. The history of this problem dates back
to the '30s when it was first pointed out by Bardeen in Ref.\cite{bardeen} . A
preliminary treatment of this work appeared in Ref.\cite{rebei}. It 
was argued in Ref.\cite{rebei} that a 
correct approximation to the internal energy of a
many-particle system up to exchange, i.e., a Hartree-Fock approximation, will
automatically include summing an infinite set of diagrams generated by the
known ``anomalous'' diagram, Fig.~\ref{anomalous}.\newline 

At low temperature, the ring diagrams treated in Ref.\cite{gellmann} are 
of order
$e^{4}$. The exchange energy is of order $e^{2}$. However at nonzero
temperature, we have another contribution to order $e^{2}$ that comes from
summing an infinite set of diagrams that were never treated before. Therefore
the new set can not be ignored in any finite temperature treatment. We show
below that taking these diagrams into account is exactly what is needed to
solve the anomaly in the exchange energy at finite temperature. The grand
partition function $Z$ of the system is given in this case by the following expression:%

\begin{equation}
Z=\exp(-\beta\Omega)=Tr\exp\left(  -\beta\left[  \left(  \mathrm{H-\mu
N}\right)  +j\varphi+\frac{1}{2}\varphi\mathrm{B\varphi+\psi^{\dag}Q\psi
}\right]  \right)  \text{ }.
\end{equation}
\newline The thermodynamical potential is a functional of $j$, $B$ and $Q$ but
it is a function of the variables $\mu,$ $T$ and $V$. To get the free energy
$F[\rho,\varphi_{c},\Delta]$ we make the following triple Legendre
transformation
\begin{equation}
F[\rho,\varphi_{c},\Delta]-\mu\,N=\Omega\lbrack j,B,Q]-j\varphi_{c}-\frac
{1}{2}\varphi_{c}B\varphi_{c}-\rho Q-\Delta B
\end{equation}
where $\rho$, $\varphi_{c}$ and $\Delta$ are defined as before, Eqs.(18-20).
We have seen that ${F}$ ($E$ at $T=0$) can be calculated perturbatively by
expanding the exponent around the classical Hartree potential in a neutral
background, i.e., with no external potential. The coefficients of this
expansion are expressible in terms of Feynman diagrams. Here in this section
we keep only the first two terms of the expansion, in other words we only
include the first diagram in Fig.~\ref{expansion}.

The equation of motion satisfied by $\rho(x,x^{\prime})$ is given by:%

\begin{equation}
{\left(  \frac{\partial}{\partial\tau}-\frac{1}{2}{\nabla}_{x}^{2}-\mu\right)
\rho(x-y)}=\delta(x-y)-e^{2}\int dz\rho(x-z)\rho(z-y)\Delta(z-x)\text{ },
\end{equation}
\newline where $\Delta(z-x)$ is the bare Coulomb potential to first
approximation. It must be pointed out again  that a full solution to
$\Delta(z-x)$ gives the expected shielded potential. An approximate solution
to the full linearized equation for $\Delta(z-x)$ was first given
in Ref.\cite{matsubara}. We must 
stress however that in what follows we use only
the zero order solution to $\Delta(z-x).$ In this case the above equation is
the Hartree-Fock approximation to the equation of motion for the one-particle
Green's function $\rho(x-y)$ at finite temperature \cite{martin}. The 
final
expression that we are after is that for the thermodynamic potential
$\Omega(\mu,T,V)$. Hence within the above approximation, the expression for
the thermodynamic potential has the following form:%

\begin{align}
\Omega & =\Omega_{0}+\Omega_{ex}+\frac{1}{2}Tr \left \{ \ln \left [ 
\delta(x-y)+e^{2}\Re(x-y) \right ] -e^{2}\Re(x-y) \right \} \nonumber\\
& -Tr \left \{ \ln \left [ \delta(x-y)-e^{2}\Im(x-y) \right ] +e^{2}
\Im(x-y) \right \}
\end{align}
where the functions $\Re(x-y)$ and $\Im(x-y)$ are given by
\begin{equation}
\Re(x-y)=\int_{0}^{\beta}d\tau\int d^{3}zA(y-z)\rho(x-y)\rho(z-x)
\end{equation}
and
\begin{equation}
\Im(x-y)=\int_{0}^{\beta}d\tau\int d^{3}zA(x-z)\rho(z-x)\rho(y-z)\: .
\end{equation}
The term $\Omega_{0}$ is the free contribution, $\Omega_{ex}$ is the usual
first order exchange term, the third term represents the usual ring diagrams
and the last term represents a new series of diagrams shown in
Fig.~\ref{anomalous}. The function $A(x-y)$ is the bare Coulomb potential
and $\rho(x-y)$ is the free one-particle Green's function at finite
temperature. It should be stressed from the above treatment that the new
diagrams are necessary for the consistency of our approximations. The first
diagram of the new set of diagrams was treated previously by Isihara and
Kojima in Ref.\cite{kojima} in a 
calculation of the specific heat. Here we do the
same calculation again but taking into account the full sequence of the new
diagrams. The method used by these authors differs from ours, however they are
both based on path integrals and both start from a partition function. In this
sense our methods are more exact than those that start from a zero temperature
treatment like Gell-Mann's treatment in Ref.\cite{gellmann2} or those 
based on a
phenomenological theory like Landau's. The authors of Ref.\cite{kraeft}
claim at the end of their paper that screening is probably not needed to have
a reasonable answer for the specific heat. Here we show that indeed this claim
is true. It is well known that the exchange energy has an odd $\ln T$ behavior
that ends up contributing a $T\ln T$ term to the specific heat. This is
obviously not reasonable for low temperatures. The exchange energy at low
temperature is given by  \cite{horowitz},\cite{isihara} :

\begin{equation}
\Omega_{ex}=-V\frac{e^{2}m^{2}}{2\pi^{2}}\left \{ \frac{2}{\pi}\mu^{2}%
-\frac{\pi}{3\beta^{2}}(1+\ln\beta\mu-\gamma)\right \} \: .\label{E2}%
\end{equation}
\newline Now we give the contribution due to the ring diagrams at finite
temperature  \cite{rebei1}. This contribution is given by%

\begin{equation}
R\;=\frac{1}{2}Tr \left \{ ln \left [\delta(x-y)+e^{2}\Re(x-y)\right ]
-e^{2}\Re(x-y) \right \} .
\end{equation}
\newline In momentum representation, this is given by%

\begin{equation}
R\;=\frac{V}{2\beta}\,\sum_{n=-\infty}^{n=\infty}\frac{4\pi}{(2\pi)^{3}}%
\int_{0}^{\infty}\,p^{2}\,dp\;\left [\ln \left \{ 1+e^{2}\Re(p,\omega
_{n})\right \}-e^{2}\Re(p,\omega_{n})\right ]
\end{equation}
\newline where p represents momentum and $\omega_{n}=\frac{(2n+1)\pi}{\beta}$,
$n=0,\pm1,\pm2,\ldots$. The function $\Re(p,\omega_{n})$ is given by:

\begin{equation}
\Re(p,\omega_{n}) \; = \; \frac{2}{(2\pi)^{3}p^{2}}\!
\int\!d\vec{q}\frac{\frac{1}
{2}(\vec{p}+\vec{q})^{2}-\frac{1}{2}\vec{q}^{2}}{ \omega_{n}^{2}+\left (
\frac{1}{2}(\vec{p}+\vec{q})^{2}-\frac{1}{2}\vec{q}^{2} \right )^{2}}
(f(\vec{q})-f(\vec{p}+\vec{q})).
\end{equation}
 
\noindent The function $f(\vec{q})$ is the Fermi-Dirac function,%

\begin{equation}
f(\vec{q})\;=\;\frac{1}{1+\exp\left [\beta(\frac{q^{2}}{2}-\mu)\right ]} .
\end{equation}\\
 Therefore the finite temperature contribution of the ring diagrams to
leading order in ${\frac{1}{\eta^{2}}}$, $\eta=\beta\mu$, and for high density
is given by
\begin{equation}
R\,=\,N(0.0622\ln r_{s}-0.142)+0.0181\,\mathit{N}(T)\;\mu\;(\frac{\alpha
r_{s}(T)}{\eta})^{2}\text{ }.
\end{equation}
\newline Here $N$ is the 
total number of particles, which is a constant. All the
other variables are defined as follows:
\begin{equation}
\alpha=(\frac{4}{9\pi})^{\frac{1}{3}},\: \: r_{s}(T)=\frac{me^{2}}
{\alpha\hbar^{2}k_{F}},\: \: \,a_{0}=\frac{\hbar^{2}}{me^{2}},\: \:
 \mathit{N}(T)=\frac{V}{\frac{4\pi}{3}r_{s}(T)^{3}a_{0}^{3}} \: .
\end{equation}

 To obtain this answer, we have used explicitly the fact that $r_{s}$
is small to simplify the calculation. It is however important  to stress that
this approximation does not affect our final result regarding the $\ln T$ 
terms in the specific heat. Obviously our result will not 
apply even to $r_{s} \approx 1$.  

\bigskip

The contribution of the new set of diagrams, Fig.~\ref{anomalous}, is given
by :

\begin{align}
\mathcal{A}  & =Tr \left \{ \text{ }\ln\left[  \text{ }\delta(x-y)-e^{2}%
\Im(x-y)\right]  +e^{2}\Im(x-y) \right \} \nonumber\\
& =\frac{2V}{\beta}{\sum_{n=-\infty}^{\infty}}\frac{4\pi k_{F}^{3}}{{\ (2\pi
)}^{3}}\int_{0}^{\infty}x^{2}dx\left[ \: \ln \left ( 1-e^{2}%
\Im(x,\omega_{n}) \right )+e^{2}\Im(x,\omega_{n})\right]  .
\end{align}
\newline Here $x=\frac{k}{k_{F}}$ is the normalized momentum and $\omega
_{n}=\frac{(2n+1)\pi}{\beta}$ for $n=0,\pm1,\pm2,...$. The function
$\Im(x,\omega_{n})$ is given by :%

\begin{equation}
\Im(x,\omega_{n})=\frac{1}{\pi^{2}}\rho(x,\omega_{n}) \int d^{3}\vec{p}%
\frac{1}{1+\exp(\eta(\frac{p^{2}}{2} -1))} \frac{1}{\mid\vec{p} -x\vec{ k_{F}%
}\mid^{2}}%
\end{equation}
\newline with%

\begin{equation}
\rho(k,\omega_{n})=\frac{i}{i\omega_{n}-(\frac{k^{2}}{2}-\mu)}%
\end{equation}
\newline and $\eta=\beta\mu$. The mass of the electron is taken to be m=1. The
calculation is carried out for large $\eta$, i.e. for low temperature. Again
here we used explicitly that $r_{s}$ is a small parameter. After some lengthy
calculations, we get the following explicit expression for $\mathcal{A}$ to
order $\frac{1}{\eta^{2}}$ :%

\begin{equation}
\mathcal{A}= \frac{2e^{2}V} {\pi^{3}} k_{f}^{4} \left \{ 
\frac{\pi^{2}} {24\eta^{2}} 
\left ( \zeta+\frac{1}{2}\ln\eta \right ) \right \} \label{E1}
\end{equation}\\
 where $\zeta$ is a constant.

There are two things to note. First there is \emph{no} zero temperature
contribution due to these diagrams, which is as it is supposed to be. Second,
the sum of the new diagrams gives an answer which is of order $e^{2} $, that
is it has the same order as the first order exchange energy even though the
terms in the sum are of order at least $e^{4}$. This is the reason we believe
that an expansion in $e^{2}$ did not pick up this new infinite set of
diagrams. It is also of interest to note that the ring diagrams add up to a
term of order $e^{4}$. Now a close comparison of the three terms shows that
the $\ln T$ term in the exchange energy gets cancelled by the one found from
the new set of diagrams. The ring diagrams contribute a temperature dependent
term of order $e^{4}$ and hence they are not needed to solve the $\ln T$-term
in the exchange. In other words screening does not play any role in this odd
behavior of the exchange energy or for that matter of the specific heat as we
show next.

The specific heat formula is:%

\begin{equation}
C_{V}\,=\,-T\frac{\partial^{2}F}{\partial T^{2}}%
\end{equation}
where the free energy $F$ is given by, $F(T,V,N)=\Omega+\mu N$. Using standard
techniques  \cite{fetter}, we calculate 
the contribution of all of the above
terms to the specific heat. To first order in $\frac{1}{\beta^{2}} $, we have
\cite{rebei1} :

\begin{equation}
F(T,V,N)\,=\,F_{0}(T,V,N)+\frac{k^{2}T^{2}}{\mu_{0}^{2}}\left(  -\frac
{20.19}{r_{s}}-0.077\ln r_{s}+0.184\right)
\end{equation}
\newline where $F_{0}$ is the free energy of a free electron gas. Therefore,
if $C_{V}^{0}$ is the specific heat for the ideal Fermi gas, we get for the
specific heat at low temperatures and high densities the expression,%

\begin{equation}
\frac{C_{V}}{C_{V}^{0}}\,=\,1.00+0.185r_{s}-r_{s}^{2}(0.0017-0.0007\ln
r_{s})\text{ }.
\end{equation}
\newline The constants $\zeta$ and $\gamma$ are approximately equal to 8.0 and
2.0, respectively. Now we turn again to the zero temperature case and
calculate the correction due to the inclusion of diagram $b$,
Fig.~\ref{expansion}, in the expansion of the energy.

\newpage

%

\section{The Homogeneous Electron Gas at Zero \\ Temperature}

\begin{figure}[ptb]
\resizebox{\textwidth}{!}
{\includegraphics[0in,0in][8in,3in]{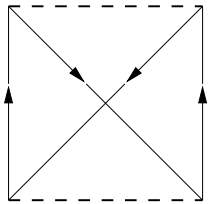}}\caption{Second order exchange
diagram in Eq.~\ref{eq:Ec}. }
\label{exch2}
\end{figure}

\begin{figure}[ptb]
\resizebox{\textwidth}{!}
{\includegraphics[0in,0in][8in,4in]{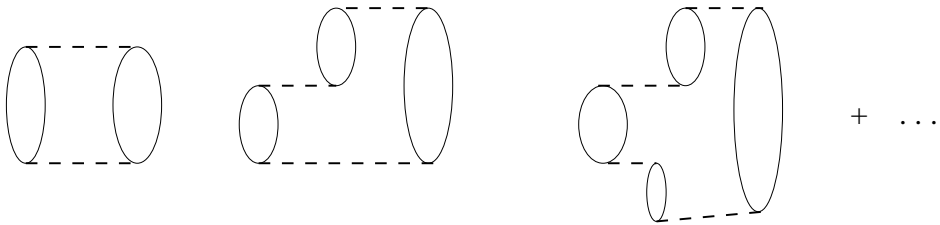}}\caption{Ring diagrams generated
by $\mathcal{H}(p)$ in Eq.~(\ref{eq:Ec}). }%
\label{ring}%
\end{figure}

\begin{figure}[ptb]
\includegraphics{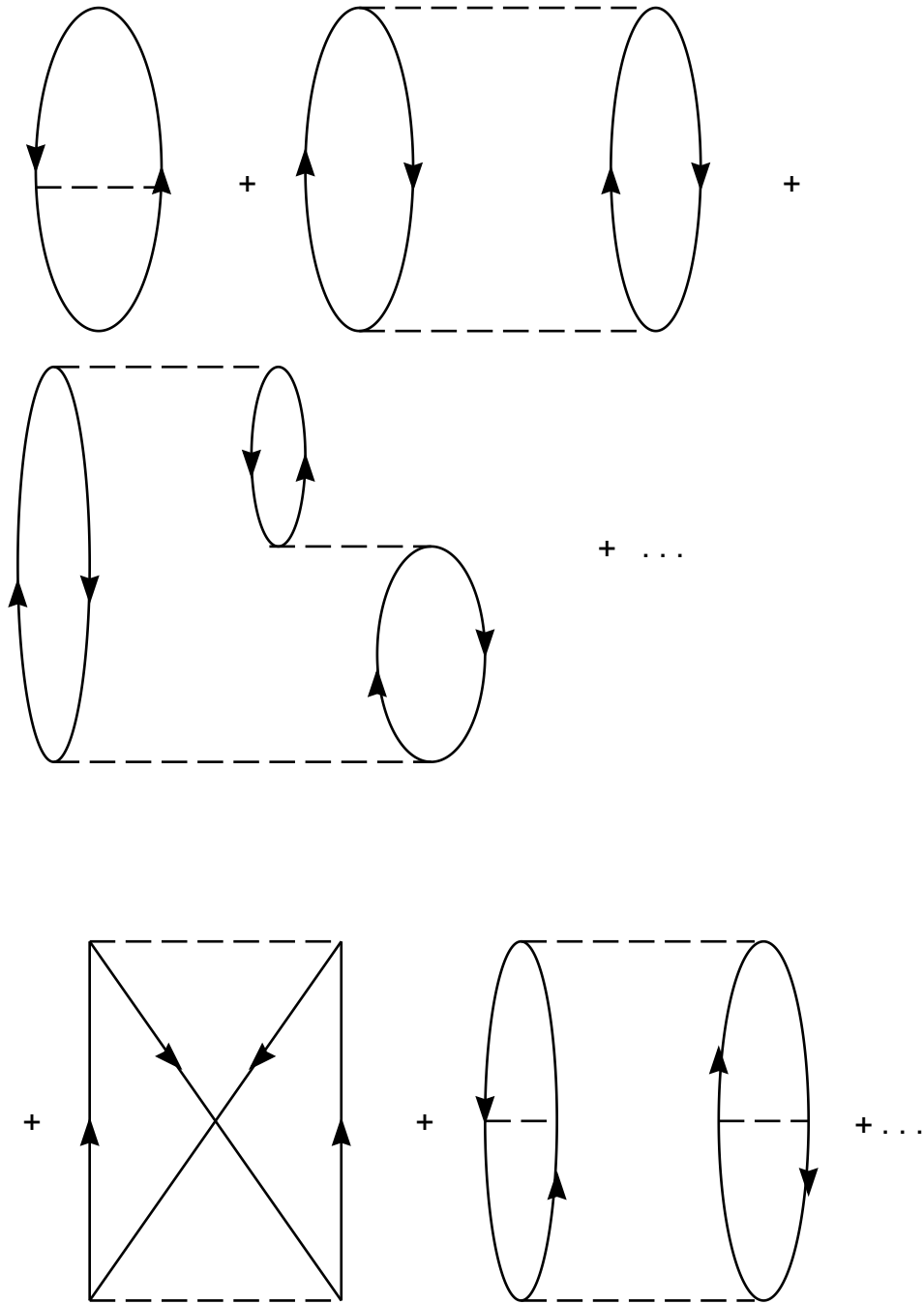}\caption{Diagrams that appear in the first ln term
in Eq.(~\ref{eq:Ec}). }%
\label{firstlog}%
\end{figure}

\begin{figure}[ptb]
\resizebox{\textwidth}{!}
{\includegraphics[0in,0in][8in,4in]{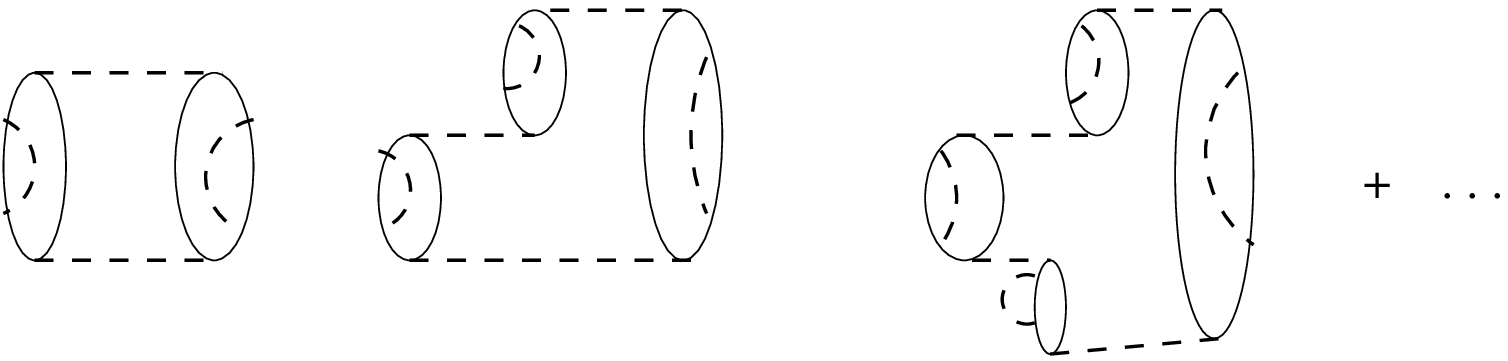}}\caption{Ring diagrams
generated by $\Pi(p)$ in Eq.~(\ref{eq:Ec}). }%
\label{ringSE}%
\end{figure}

\begin{figure}[ptb]
\includegraphics{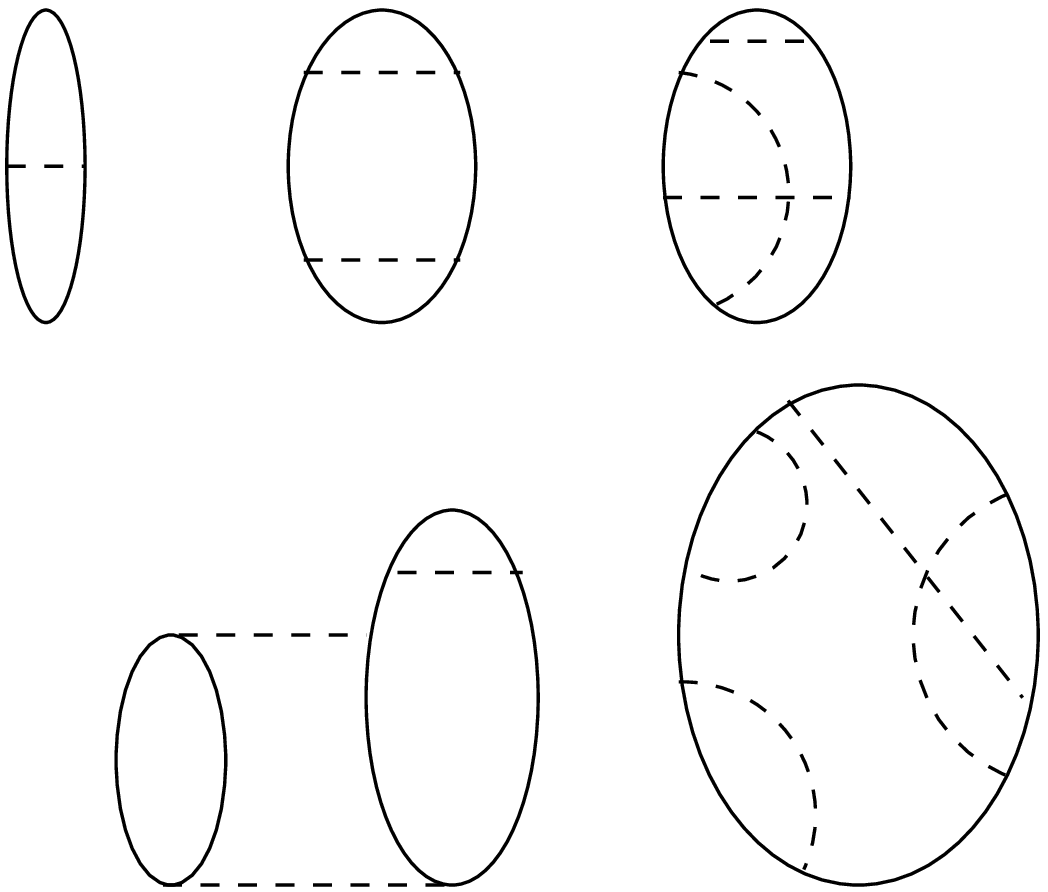}\caption{Diagrams that appear in the second ln term
in Eq.(~\ref{eq:Ec}).}%
\label{secondlog}%
\end{figure}

\begin{figure}[ptb]
\resizebox{\textwidth}{!}
{\includegraphics[0in,0in][8in,7in]{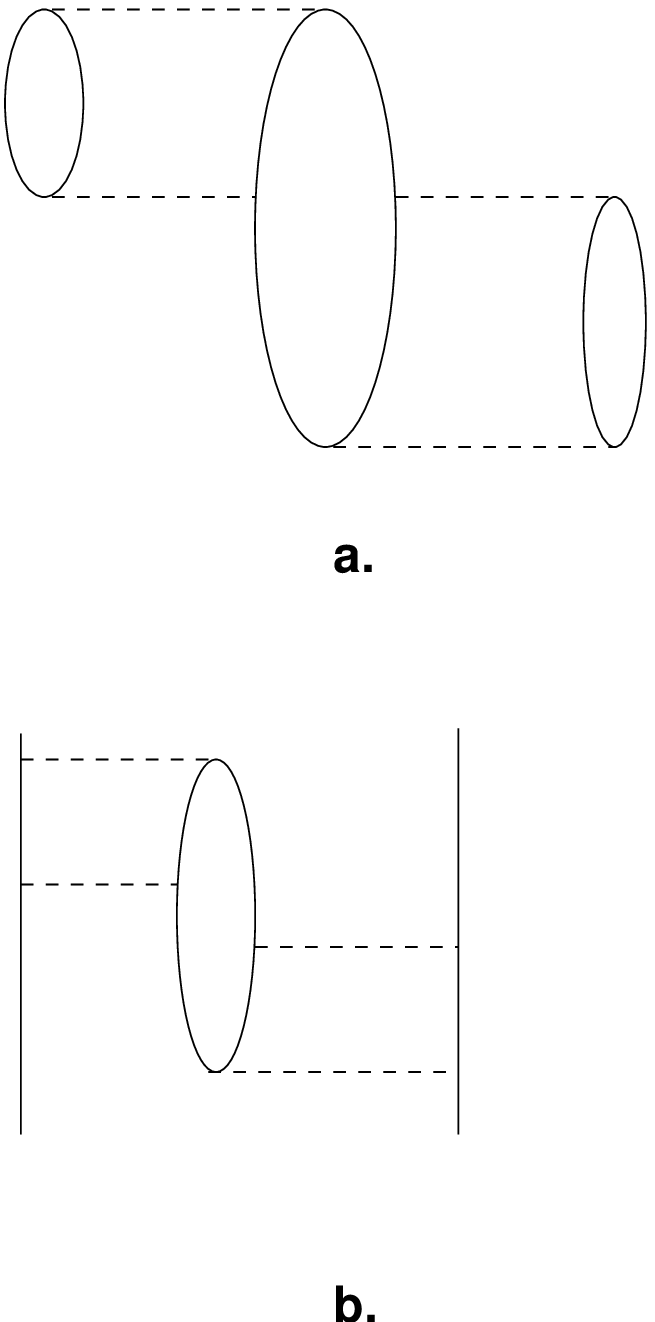}}\caption{A term of order
$e^{4}$ that is due to $\Sigma_{1}$. Both representations are equivalent. 
Initially two particles interact with each other with one of them going back to its
initial state while the other one interacts with a third particle before
both returning to their corresponding initial states.  }
\label{sigma1}%
\end{figure}

\begin{figure}[ptb]
\resizebox{\textwidth}{!}
{\includegraphics[0in,0in][8in,9in]{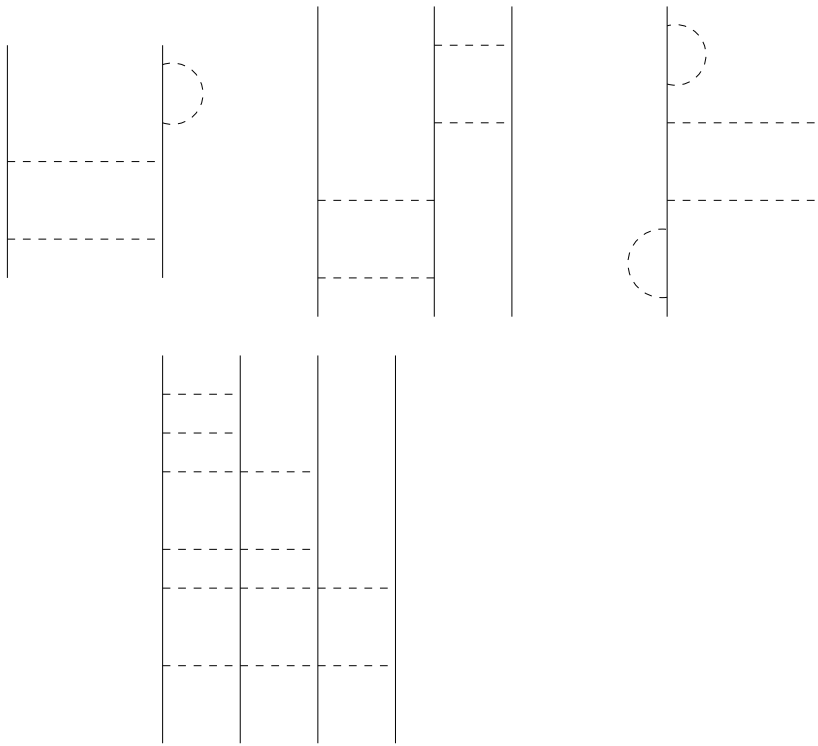}}\caption{ Some of the ladder
diagrams due to $\Sigma_{1}$.}%
\label{ladder3}%
\end{figure}

\begin{figure}[ptb]
\resizebox{\textwidth}{!}
{\includegraphics[0in,0in][8in,8in]{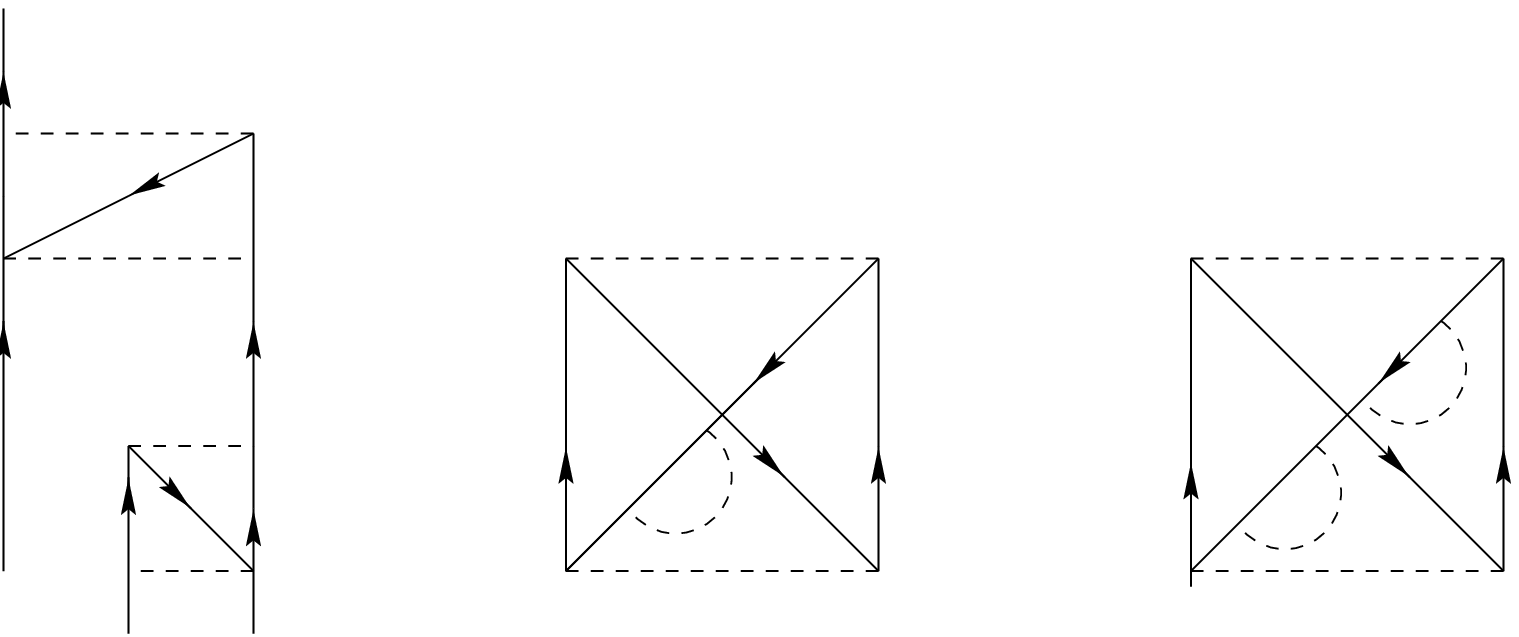}}\caption{Few terms that
appear due to $\Sigma_{2}$.}%
\label{sigma2}%
\end{figure}

\begin{figure}[ptb]
\resizebox{\textwidth}{!}
{\includegraphics[0in,0in][8in,7in]{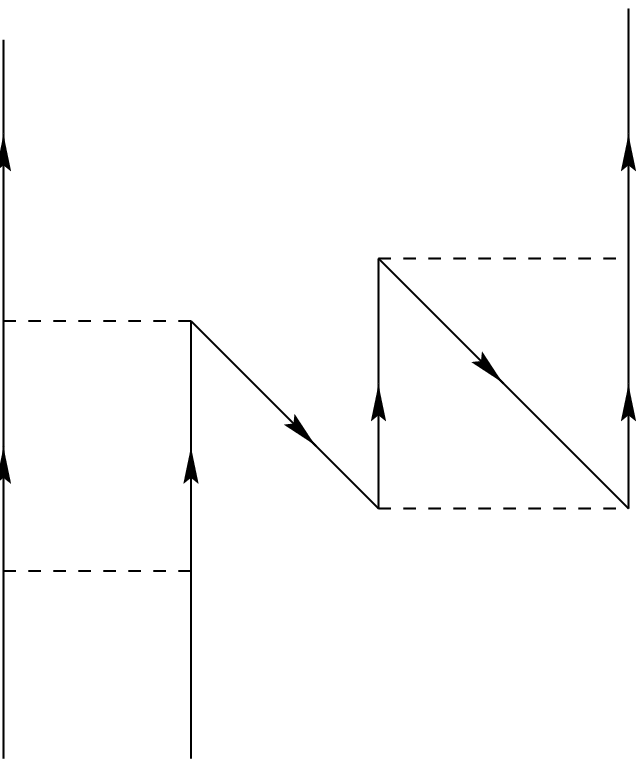}}\caption{A term that is
both due to $\Sigma_{1}$ and $\Sigma_{2}$. Here three particles are 
interacting pairwise with two of the three particles exchanging states.}%
\label{crossTerm}%
\end{figure}

\begin{figure}[ptb]
\resizebox{\textwidth}{!}
{\includegraphics[0in,0in][8in,6in]{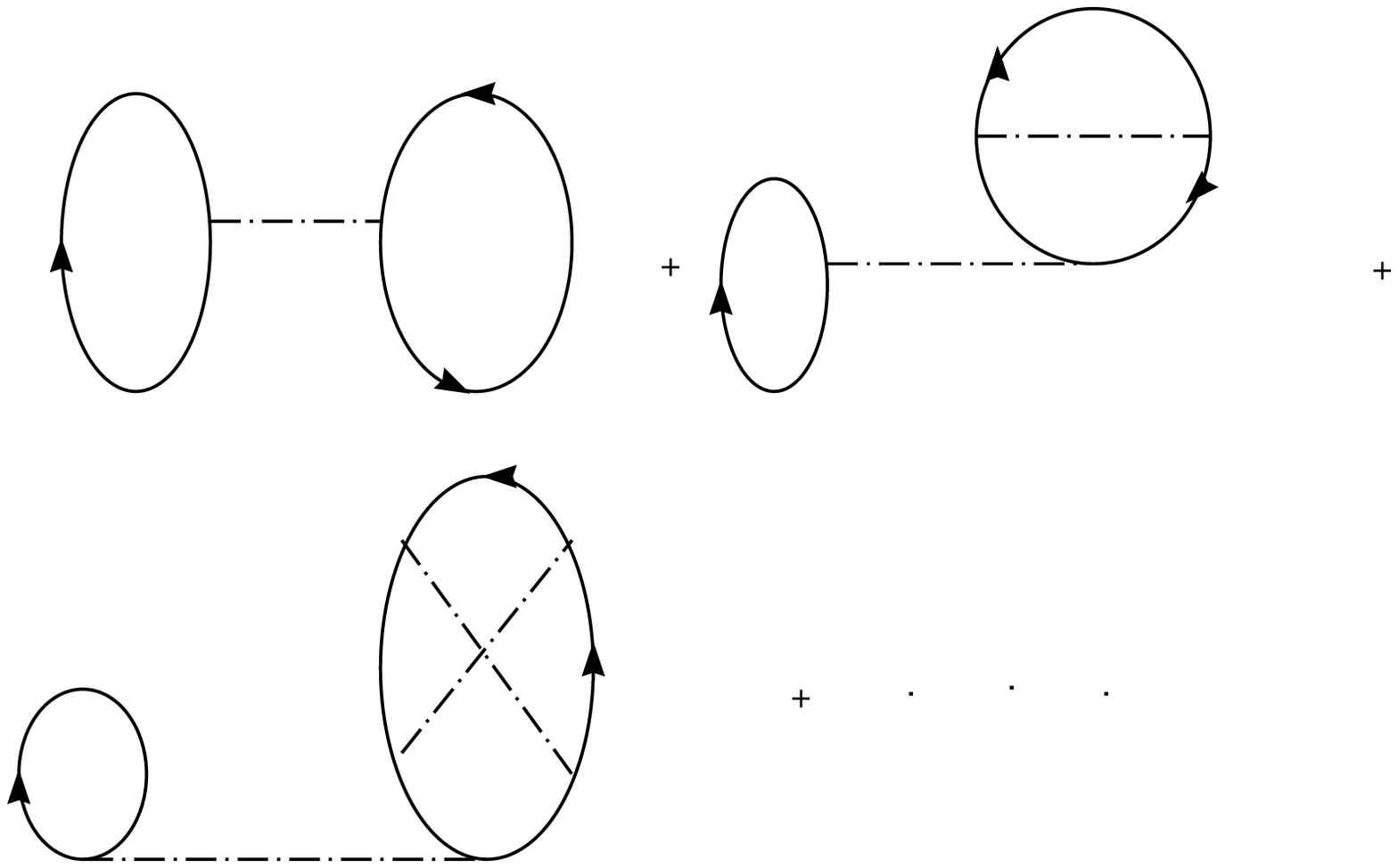}}\caption{Coulomb interactions
that must be taken into account in inhomogeneous media. These diagrams 
vanish in the homogeneous case.}%
\label{coul}%
\end{figure}

In this section we apply the main result of section 4, the expression for the
energy, to the homogeneous case at zero temperature.  We assume that there is
a background of positive charge of equal magnitude to the average density of
the electron gas. Hence the system is neutral. The literature on this problem
of the calculation of the correlation energy is huge. The most complete
treatment, so far as we are aware, was given by Bishop and Luhrmann
 and it was restricted to zero 
temperature effects \cite{bishop} . Since the
system is homogeneous, the final expression for the energy will be given in
momentum space. The Green's function that will be used in the following is the
solution to the first iteration of the nonlinear equation satisfied by
$\rho(x,y)$. Since there is no external potential, the input Green's function
is that of a free electron gas. The energy expression is given in imaginary
time, hence we use the following expression for the free Green's function%

\begin{equation}
\rho_{0}(\omega,k)=\frac{1}{-i\omega+\frac{1}{2m}k^{2}-\mu} \: \: .
\end{equation}
\newline In this section and the next, the electron Green's function
definition differs by a factor of i from the one given before. In terms of
$\rho_{0}(x,y)$, the ground state energy is given by the following, omitting
the subscript 0 for simplicity,%

\[
-T\;E_{gs}=\int d^{4}x\,\left(  \frac{1}{2}i\nabla^{2}\rho(x,y)\right)
_{x=y}\text{ }
\]
\[
\text{\ }+\frac{e^{4}}{4}\int dx\,dz\,\delta(x-z)\;\mathcal{M}(x,z)+\frac
{e^{2}}{2}\int dx\,dy\;\rho(x,y)\rho(y,x)A(x,y)
\]
\[
-\frac{1}{2}Tr_{(x,z)}\text{ \ }\ln\text{ }\left[  \delta(x,z)-e^{2}%
\mathcal{H}(x,z)+\text{ }e^{4}\,\mathcal{M}(x,z)-2\text{ }e^{4}\Pi(x,z)\text{
}\right]  \text{ \ }
\]
\[
+Tr_{(x,z)}\text{ \ }\ln\left[  \delta(x,z)-e^{2}\mathcal{K}(x,z)-e^{4}%
\;\Sigma(x,z)\text{ }\right]  \text{
\ \ \ \ \ \ \ \ \ \ \ \ \ \ \ \ \ \ \ \ \ \ \ \ \ \ \ \ \ \ }
\]
\begin{equation}
+\int dx\,dz\,\delta(x-z)\;\left\{  \frac{e^{2}}{2}\mathcal{H}(x,z)-\frac
{e^{4}}{2}\mathcal{M}(x,z)+e^{4}\Sigma(x,z)+e^{4}\Pi(x,z)\right\}  ,
\end{equation}
where
\begin{equation}
\mathcal{M}(x,z)=\int dy\,dx_{1}\,dx_{2}\,\rho(y,x_{2})\rho(x_{2}%
,x)\rho(x,x_{1})\rho(x_{1},y)A(x_{1},x_{2})A(y,z)\text{ },
\end{equation}%
\begin{equation}
\mathcal{H}(x,z)=\int dy\,A(z,y)D(x,y)\text{ },
\end{equation}%
\begin{equation}
\Pi(x,z)=\int dy\,dx_{1}\,dx_{2}\,\rho(y,x)\rho(x,x_{1})\rho(x_{1},x_{2}%
)\rho(x_{2},y)A(x_{1},x_{2})A(y,z)\text{ },
\end{equation}%
\begin{equation}
\mathcal{K}(x,z)=\int dy\,A(x,y)\rho(x,y)\rho(y,z)\text{ },
\end{equation}%
\begin{equation}
\Sigma(x,z)=\Sigma_{1}(x,z)-\Sigma_{2}(x,z)\text{ },
\end{equation}%
\begin{equation}
\Sigma_{1}(x,z)=\int dy\,dx_{1}\,dx_{2}\,\rho(x,y)\rho(y,z)\rho(x_{1}%
,x_{2})\rho(x_{2},x_{1})A(x,x_{1})A(x_{2},y)\text{ },
\end{equation}%
\begin{equation}
\Sigma_{2}(x,z)=\int dy\,dx_{1}\,dx_{2}\,\rho(x,x_{2})\rho(x_{2},x_{1}%
)\rho(x_{1},y)\rho(y,z)A(x_{1},x)A(x_{2},y)\text{ }.
\end{equation}
\newline The above equation, Eq.(82), is simply Eq.(59) taking into account of
the fact that $\rho(x,x)$ gets canceled by an equal and opposite charge.
Because of the transitional invariance of the system, we can write a simple
expression for the correlation energy in the momentum representation,%

\begin{align}
E_{c}  & =\frac{1}{4}e^{4}V\int\prod_{i=1}^{3}\frac{d^{3}p_{i}}{(2\pi)^{3}%
}\frac{d\omega_{i}}{2\pi}A(p_{1}-p_{2})A(p_{2}-p_{3})\rho(p_{1})\rho
(p_{2})\rho(p_{3})\rho(p_{1}-p_{2}+p_{3})\nonumber\\
& +\frac{1}{2}V\int\frac{d^{3}p}{(2\pi)^{3}}\frac{d\omega}{2\pi} \left \{
\ln \left [ 1+e^{2}\mathcal{H}(p)+e^{4}\mathcal{M}(p)-2e^{4}\Pi
(p)\right ]-e^{2}\mathcal{H}(p)-e^{4}\mathcal{M}(p)+2e^{4}\Pi(p)\right \}
\nonumber\\
& -V\int\frac{d^{3}p}{(2\pi)^{3}}\frac{d\omega}{2\pi}\left\{  \;\ln
\left [1-e^{2}\mathcal{K}(p)-e^{4}\Sigma(p) \right ] +e^{2}\mathcal{K}
(p)+e^{4}\Sigma(p)\;\right\}  \;.\label{eq:Ec}
\end{align}
\newline Since the ring diagrams are part of $E_{c}$, this will be the first
test to see if our expansion gives the correct leading result. First we show
how this equation follows from Eq.(82). We start by letting%

\begin{equation}
\mathcal{H}(x,z)=-\int\,dy\: D(x,y)A(y,z)
\end{equation}
or, since the system is homogeneous, we can write instead%

\begin{equation}
\mathcal{H}(x-z)=\int dy\,D(x-y)A(y-z).
\end{equation}
\newline Next, we rewrite $D(x-y)$ and $A(x-z)$ in terms of their
corresponding Fourier transforms, i.e.,%

\begin{equation}
D(x-y)=\int\frac{dk}{(2\pi)^{4}}\;\exp[ik\cdot(x-y)]D(k)
\end{equation}
and
\begin{equation}
A(y-z)=\int\frac{dq}{(2\pi)^{4}}\;\exp[iq\cdot(y-z)]A(q)\text{ }.
\end{equation}
\newline Hence, by convolution we have,%

\begin{equation}
\mathcal{H}(p)\,=\,(A\,\star\,D)\;(p)\text{ },
\end{equation}
\newline however since
\begin{equation}
D(p)=\int\frac{dq}{(2\pi)^{4}}\text{ }\rho(p+q)\,\rho(q),
\end{equation}
\newline it follows then that,%

\begin{equation}
\mathcal{H}(p)\,=\, A(p)\,\int\frac{dq}{(2\pi)^{4}} \, \rho(p+q) \rho(q) .
\end{equation}
\newline After using the expressions for the propagators, we end up with the
following expression for $\mathcal{H}(p)$%

\begin{equation}
\mathcal{H}(p)=\frac{4\pi}{p^{2}}\!\int\!\frac{dq}{(2\pi)^{4}}\,\frac
{1}{-i\bar{\omega}+\frac{1}{2}\,{\vec{q}}^{2}-\mu} \; \frac{1}{-i(\bar{\omega
}+\omega)+\frac{1}{2}\,(\vec{p}+\vec{q})^{2}-\mu}%
\end{equation}
\newline with $q=(\bar{\omega},\vec{q})$. Integrating first over $\bar{\omega
}$ in the complex plane, we find that%

\begin{equation}
\mathcal{H}(p)=-2e^{2}\frac{4\pi}{p^{2}}\!\!\int\!\!\frac{d\vec{q}}{(2\pi
)^{3}}\frac{\left(  \frac{1}{2}(\vec{p}+\vec{q})^{2}-\frac{1}{2}\vec{q}%
^{2}\right)  \left(  \Theta(\mu-\frac{1}{2}\vec{q}^{2})-\Theta(\mu-\frac{1}%
{2}(\vec{p}+\vec{q})^{2})\right)  }{\omega^{2}+\left(  \frac{1}{2}(\vec
{p}+\vec{q})^{2}-\frac{1}{2}\vec{q}^{2}\right)  ^{2}}\label{eq:AD}%
\end{equation}
\newline where we added a factor of $2$ to account for spin. This term
$\mathcal{H}(p) $ will prove to be all that is needed to reproduce the known
RPA result. All other terms that appear in the energy expression are new
additions to the correlation. Before showing this explicitly , we give the
expressions for the remaining terms $\mathcal{K}(p)$,  $\mathcal{M}(p)$, 
$\Pi(p)$,  $\Sigma_{1}(p)$,  $\Sigma_{2}(p)$ and $\Sigma(p)$.%

\begin{equation}
\mathcal{K}(p)=\rho(p)\int\frac{dq}{(2\pi)^{4}}\text{ }\rho(p+q)\text{ }A(q)
\end{equation}%
\begin{equation}
\mathcal{M}(p)=A(p)\int\frac{dk_{1}}{(2\pi)^{4}}\frac{dk_{2}}{(2\pi)^{4}}%
\rho(k_{1})\rho(k_{1}+p)\rho(k_{2})\rho(k_{2}-p)A(k_{1}-k_{2}+p)\text{ },
\end{equation}%
\begin{equation}
\Pi(p)=A(p)\int\frac{dk_{1}}{(2\pi)^{4}}\frac{dq}{(2\pi)^{4}}\rho(k_{1}%
)\rho(k_{1}-p)\rho(k_{1}-p-q)A(q)\text{ },
\end{equation}%
\begin{equation}
\Sigma_{1}(p)=\rho(p)\int\frac{d^{4}p_{1}}{(2\pi)^{4}}\frac{d^{4}p_{2}}%
{(2\pi)^{4}}\,\rho(p_{1})\rho(p_{2})\rho(p_{2}+p_{1}-p)A(p_{2}-p)A(p-p_{1}%
)\text{ },
\end{equation}%
\begin{equation}
\Sigma_{2}(p)=\rho(p)\int\frac{d^{4}p_{1}}{(2\pi)^{4}}\frac{d^{4}p_{2}}%
{(2\pi)^{4}}\,A^{2}(p_{1})\rho(p-p_{1})\rho(p_{1}+p_{2})\rho(p_{2})
\end{equation}
and
\begin{equation}
\Sigma(p)=\rho(p)\int\frac{d^{4}k_{1}}{(2\pi)^{4}}\frac{d^{4}k_{2}}{(2\pi
)^{4}}\rho(k_{1})\rho(k_{2})\rho(k_{2}+k_{1}-p)\left[  A(k_{1}-p)^{2}%
-A(p-k_{2})A(p-k_{1})\right]  \text{ }.
\end{equation}
\bigskip The Fourier transform $A(p)$ is given as usual by%

\begin{equation}
A(\,p\,)\;=\;\frac{4\,\pi}{|\,\vec{p}\,|^{2}}\: .
\end{equation}
\newline The terms in Eq.~(\ref{eq:Ec}) have the following meaning after we
expand the $\ln$-terms. The first term represents the second order exchange
term, Fig.~\ref{exch2}.
 The term $\mathcal{H}(p)$ is responsible for
generating the ring diagrams, Fig.~\ref{ring}.
 The term $\mathcal{M}(p)$ is
responsible for generating some of the ring diagrams with exchange. The cross
terms give the remaining ring diagrams with exchange, Fig.~\ref{firstlog}.
The $\Pi(p)$ term generates ring diagrams with a self-energy insertion in each
ring, fig.~\ref{ringSE}.
 Hence the terms in the first $ln$ should contribute
the most. Some of the terms that appear upon expansion of the
second $\ln$ term are shown in Fig.~\ref{secondlog}.
 The term $\mathcal{K}(p)$ generates
the so called `anomalous' diagrams, Fig.~\ref{anomalous}. The term
$\Sigma_{1}(p)$ generates ladder terms like those in Figs.~\ref{sigma1} and
~\ref{ladder3}.
$\Sigma_{2}(p)$ generates terms like those in
Fig.~\ref{sigma2}.
Cross terms of the last two terms are shown in
Fig.~\ref{crossTerm}.
Diagrams that appear in Fig.~\ref{coul} are important in a nonhomogeneous
medium and must be accounted for since they no longer vanish. They arise
whenever we have a $\rho(x,x)$ term in the expression for energy, Eq.(59). 
 Normalizing the momentum with respect to $k_{F}$, the
Fermi momentum, the explicit expression for the correlation energy per
particle in Rydbergs (Ryd.) has the following form,

\begin{eqnarray}
 \lefteqn{ \frac{E_{c}}{N}   =\frac{3}{16\pi^{5}}
\int d^{3}q\,d^{3}p_{1}\,d^{3}
p_{2}\frac{\theta(1-p_{2})\theta(1-p_{1})
\theta(E_{q_{1}}-1)\theta(E_{q_{2}}-1) }{q^{2} (q+p_{_{1}}-p_{_{2}} )^{2}
( q^{2}+q\cdot p_{1}-q\cdot p_{2} ) } } \nonumber\\
& & \mbox{}+ \frac{3}{4\pi\alpha^{2}r_{s}^{2}} \int_{-\infty}^{\infty}
dx\int_{0}^{\infty} dy\: y^{2} \left \{ \ln \left [ 
1+\mathcal{H}(x,y)+\mathcal{M}
(x,y)-2\Pi(x,y) \right ] \right . \nonumber\\
& &  \left . -\mathcal{H}(x,y)-
\mathcal{M}(x,y)+2\Pi(x,y) \right \} \nonumber\\
& & -\frac{3}{2\pi\alpha^{2}r_{s}^{2}} \!\int_{-\infty}^{\infty}\!dx\int
_{0}^{\infty}\!dy\:y^{2} \left \{ \ln \left [ 1+\mathcal{K}
(x,y)+\Sigma(x,y)\right ] -\mathcal{K}(x,y)-\Sigma(x,y) \right \},
\label{eq:Ecp}
\end{eqnarray}
where we have set $E_{q_{1}}=|\vec{q}+\vec{p}_{1}|$ and $E_{q_{2}}=|\vec
{q}-\vec{p}_{_{2}}|$.\newline For convenience, we set
\begin{equation}
\frac{E_{c}}{N}=I_{1}\,+\,I_{2}\,+\,I_{3}%
\end{equation}
\newline where $I_{1}$, $I_{2}$, and $I_{3}$ are the first, the second and the
third integrals, respectively.  Expressions for few of the functions that 
appear above are given below. 
The functions $\mathcal{K}(x,y)$, 
$\mathcal{H}(x,y)$ and
$\Sigma_{1}(x,y)$ are given by,
\begin{equation}
\mathcal{K}(x,y)\,=\,\frac{4\alpha r_{s}g(y)}{\pi(-ix+y^{2}-1)} ,
\end{equation}

\begin{equation}
\mathcal{H}(x,y)\,=\,-e^{2}\,A(x,y)D(x,y)
\end{equation}

\medskip or more explicitly, we have
\begin{align}
\mathcal{H}(x,y)  & =\frac{2\alpha r_{s}}{\pi y^{2}}\left[  1+\frac{1}{8y^{3}%
}\left(  x^{2}+4y^{2}(1-\frac{y}{2})(1+\frac{y}{2})\right)  \ln\left(
\frac{(\frac{x}{2})^{2}+y^{2}(1+\frac{y}{2})^{2}}{(\frac{x}{2})^{2}%
+y^{2}(1-\frac{y}{2})^{2}}\right)  \right. \nonumber\\
& \left.  -\frac{x}{2y}\left(  \arctan\left(  \frac{y(\frac{y}{2}+1)}{\frac
{x}{2}}\right)  -\arctan\left(  \frac{y(\frac{y}{2}-1)}{\frac{x}{2}}\right)
\right)  \text{ }\right]
\end{align}
\newline and

\begin{equation}
\Sigma_{1}(x,y)=\frac{16\alpha^{2}r_{s}^{2}}{(2\pi)^{4}}\!\int\!d^{3}%
k_{1}d^{3}k_{2}\text{ }g(x,y,k_{1},k_{2})
\end{equation}
where
\[
g(x,y,k_{1},k_{2})=\frac{1}{|\vec{y}-\vec{k}_{1}|^{2}}\frac{1}{|\vec{y}%
-\vec{k}_{2}|^{2}}\times\frac{\left(  \theta(1-k_{1})-\theta(1-k_{12})\right)
\left(  \theta(k_{1}-E^{\prime})-\theta(1-k_{2})\right)  }{(-ix+y^{2}%
-1)(ix-k_{2}^{2}-k_{1}^{2}+E^{\prime}-1)}
\]
\newline $\theta(x)$ is the step function. $E^{\prime}$, $g(y)$ and $\alpha$
are given by
\begin{equation}
E^{\prime}=|\vec{k}_{1}+\vec{k}_{2}-\vec{y}|^{2},
\end{equation}%
\begin{equation}
\alpha=\left(  \frac{4}{9\pi}\right)  ^{\frac{1}{3}},
\end{equation}
and%

\begin{equation}
g(y)=1+\frac{1-y^{2}}{2y}\ln\left|  \frac{1+y}{1-y}\right|  \text{ }.
\end{equation}
\newline The Gell-Mann-Brueckner term can be isolated from the full expression
for the correlation, Eq.(~\ref{eq:Ecp}), by rewriting the second integral,
$I_{2} $, in the following way:%

\begin{equation}
I_{2}=I_{2}^{ring}+I_{2}^{exch}+I_{2}^{seRing}%
\end{equation}
where
\begin{align}
I_{2}^{ring}  & =\frac{3}{4\pi}\frac{1}{r_{s}^{2}\alpha^{2}}\int_{-\infty
}^{+\infty}dx\int_{0}^{\infty}dy\,\: y^{2} \left \{ \ln \left [
1+\mathcal{H}(x,y) \right ] -\mathcal{H}(x,y) \right \} \:
\label{eq:ring}\\
I_{2}^{exch}  & =\frac{3}{4\pi}\frac{1}{r_{s}^{2}\alpha^{2}}\int_{-\infty
}^{\infty}dx\int_{0}^{\infty}dy\: y^{2}\left \{  \ln \left[  1+\frac
{\mathcal{M}(x,y)}{1+\mathcal{H}(x,y)} \right]  -\mathcal{M}(x,y) \right\}.
\label{eq:exch}%
\end{align}
\newline $I_{2}^{seRing}$ is what is left of $I_{2}-I_{2}^{ring}-I_{2}^{exch}%
$. The term $I_{2}^{ring}$ is indeed the full expression for the RPA term
$\frac{1}{N}E_{c}^{RPA}.$ By taking the limit $r_{s}\rightarrow0$ and $y<1$,
it reduces to the Gell-Mann-Brueckner result  \cite{gellmann}:

\begin{equation}
\frac{E_{c}^{RPA}}{N}\approx0.0622\ln r_{s}-0.142+\ldots
\end{equation}
\medskip as we show next. Going back to the expression for $\mathcal{H}(p)$,
Eq.(~\ref{eq:AD}), we have%

\begin{equation}
I_{2}^{ring}=\frac{1}{2}\int\frac{d^{3}k}{(2\pi)^{3}}\int\frac{d\omega}{2\pi
} \left [ \ln \left (  1+\mathcal{H}(k,\omega) \right )  -\mathcal{H}
(k,\omega) \right ]
\end{equation}
\newline with $\mathcal{H}(k,\omega)$ defined to be
\begin{equation}
\mathcal{H}(k,\omega)=2e^{2}\frac{4\pi}{k^{2}}\!\int\!\frac{d^{3}p}{(2\pi
)^{3}}\,\frac{\left(  \frac{1}{2}(\vec{p}+\vec{k})^{2}-\frac{1}{2}\vec{p}%
^{2}\right)  }{\omega^{2}+\left(  \frac{1}{2}(\vec{p}+\vec{k})^{2}-\frac{1}%
{2}\vec{p}^{2}\right)  }\left(  \theta(\mu-\frac{p^{2}}{2})-\theta(\mu
-\frac{1}{2}(\vec{p}+\vec{k})^{2})\right)  \text{ }.
\end{equation}
\medskip Now we set,
\begin{equation}
\mathcal{H}(k,\omega)=\frac{2e^{2}}{(2\pi)^{3}}\,\frac{4\pi}{k^{2}}
 \left ( H_{1}(k,\omega)+H_{2}(k,\omega) \right )
\end{equation}
\newline Terms other than $I_{2}^{ring}$ in Eq.(~\ref{eq:Ecp}) provide new
correlations to the RPA-term. The first term in Eq.(~\ref{eq:Ecp} ), as we
mentioned above, is the second order exchange term and has been evaluated
exactly in  Ref.\cite{onsager},

\begin{align}
E_{c}^{2exch}  & =\frac{1}{3}\ln2-\frac{3}{2\pi^{2}}\zeta(3)\\
& \simeq0.04836\text{ \ }Ryd.\nonumber
\end{align}
\medskip To compare our expansion to others we calculate the term that is
equivalent to the RPA calculation. We start by evaluating the integral $H_{1}
$. After integrating the angular variables, we have%

\begin{equation}
H_{1}(\omega,k)=\pi\int_{0}^{k_{F}}\frac{p\,dp}{k}\,\ln\left[  \;\frac
{\omega^{2}+(pk+\frac{1}{2}k^{2})^{2}}{\omega^{2}+(-pk+\frac{1}{2}k^{2})^{2}%
}\;\right]  \text{ }.
\end{equation}
\newline This integral can be easily performed over p. We get after that%

\begin{align}
H_{1}(\omega,k)  & =\frac{\pi}{k^{3}} \left \{  \: \frac{1}{2}
\ln \left [ \;\omega^{2}+k^{2}(k_{F}+\frac{k}{2})^{2} \right ]
 \left [ \omega^{2}+k^{2}(k_{F}-\frac{k}{2})(k_{F}+\frac{k}{2}) \right ]
 \right. \nonumber\\
& - \frac{1}{2} \ln \left [ \omega^{2}+k^{2}(-k_{F}+ \frac
{ k}{2} )^{2} \right ] \, \left [ \omega^{2}-k^{2}(k_{F}
+\frac{k}{2})(-k_{F}+\frac{k}{2}) \right ] \nonumber\\
& \left.  -k^{2} \omega \left \lbrack \arctan\left(  \frac{k(k_{F}
+\frac{ k}{2})}{\omega} \right)  -\arctan \left(  \frac{k(-k_{F}
+\frac{ k}{2})}{\omega}\right)  \right ] +k_{F}k^{3} \right\}  .
\end{align}
\newline The second integral $H_{2}$ in $\mathcal{H}(\omega,k)$ is seen to be
simply obtained from $H_{1}(\omega,k)$ by replacing $\vec{k}$ by $-\vec{k}$.
Therefore we have%

\begin{equation}
\mathcal{H}(\omega,k)\, = \, 2 H_{1}( \omega, k )
\end{equation}
\newline If we set ${y=\frac{k}{k_{F}}}$ and ${x=\frac{\omega}{\mu}}$, we get%

\begin{align}
\mathcal{H}(x,y)  & =\frac{2\alpha r_{s}}{\pi y^{2}}\left\{  1+\frac
{x^{2}+4y^{2}(1-\frac{y^{2}}{4})}{(2y)^{3}}\ln\left[  \frac{x^{2}
+4y^{2}(1+\frac{ y}{2})^{2}}{x^{2}+4y^{2}(1-\frac{ y}{2})^{2}
}\right]  \right. \nonumber\\
& \mbox{}\left.  -\frac{x}{2y}\left[  \arctan\left(  \frac{2y(1+\frac
{ y}{2})}{x}\right)  +\arctan\left(  \frac{2y(1-\frac{ y}{2}%
)}{x}\right)  \right]  \right\}
\end{align}
\newline where $r_{s}=\frac{e^{2}}{\alpha\,k_{F}}$. Hence the ring diagrams'
contribution to the energy is given by%

\begin{equation}
E_{c}^{RPA}=\frac{3N}{4\pi\alpha^{2}r_{s}^{2}}\int_{0}^{\infty}y^{2}%
\,dy\int_{0}^{\infty}dx\left[  \ln[\text{ }1+\mathcal{H}(x,y)\text{
}]-\mathcal{H}(x,y)\right]  ,
\end{equation}
\newline where $N$ is the total number of electrons. Now, we notice that if we
make the following substitutions:%

\begin{align}
y\,  & =\,y^{\prime}\nonumber\\
x\,  & =\,2yx^{\prime}%
\end{align}
\medskip we recover exactly the same expression for the ring diagrams as
obtained by Bishop and Luhrmann in Ref.\cite{bishop} using a totally different
expansion. This helps to validate our original expansion in $\hbar$ and later
in iterating the equations of motion in terms of $e^{2}$. We stress that this
expansion is valid for both small and large $r_{s}$.

From the above analysis, we see explicitly that the method of effective action
amounts to including another infinite set of diagrams besides the usual ring
diagrams. One subset of the diagrams added is the ring diagrams that allow
exchange in them, 
Fig.~\ref{firstlog}. On physical grounds these are
expected to be the next important ones that must be summed up. It is also
easily seen from the above that all second order diagrams are included in the
expansion with the right symmetry factors. This shows that our original
expansion in $\hbar$ is indeed meaningful. One last thing to note about the
diagrams in Fig.~\ref{secondlog} is that they include the ``anomalous''
diagrams  \cite{luttinger}. We have already dealt with these diagrams in the
previous section. From a nonzero temperature calculation, we were able to show
that these diagrams give a zero contribution at zero temperature simply
because they violate Fermi statistics.  From 
Eq.(107), the contribution of these diagrams involves integrating $\ln \left [
1 + \mathcal{K}(x,y) \right ] - \mathcal{K}(x,y)$ over all $x$ in the 
complex plane. Since $\mathcal{K}(x,y)$ has a simple pole, Eq.(109), the above 
integrand ends up having poles of order two and higher and hence their zero
contribution at zero temperature. However these diagrams become essential at
non-zero temperature where the above constraint of Fermi statistic is 
no longer an issue. In the following section  we calculate 
the contribution of the function
$\mathcal{M}(x,y)$ to the correlation energy. 

\newpage

%

\section{The Inclusion of Second Order Exchange Effects \\ in Ring Diagrams}

\begin{figure}[ptb]
\resizebox{\textwidth}{!}
{\includegraphics[0in,0in][8in,8in]{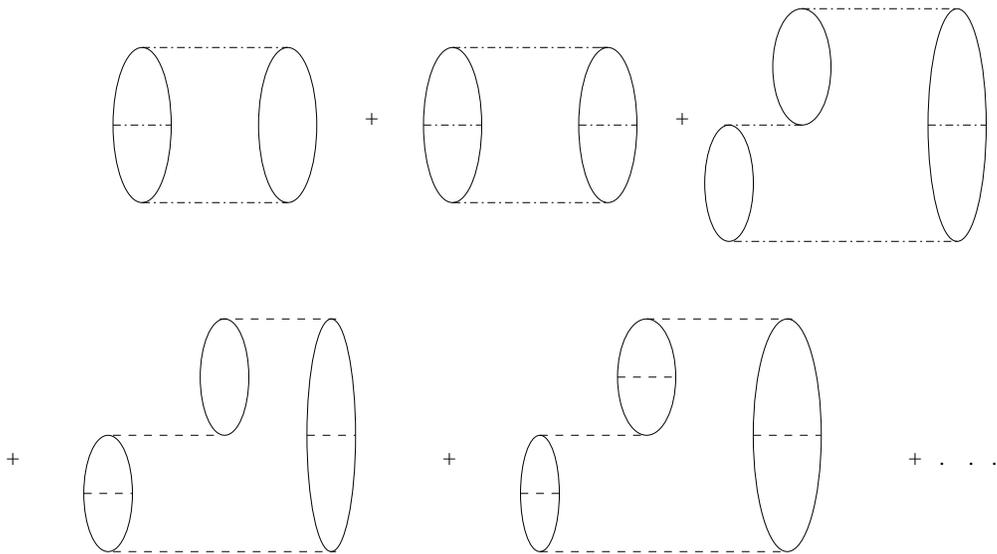}}\caption{Ring diagrams
with exchange.}%
\label{RingExchange}%
\end{figure}

\begin{figure}
\resizebox{\textwidth}{!}
{\includegraphics[0in,0in][8in,8in]{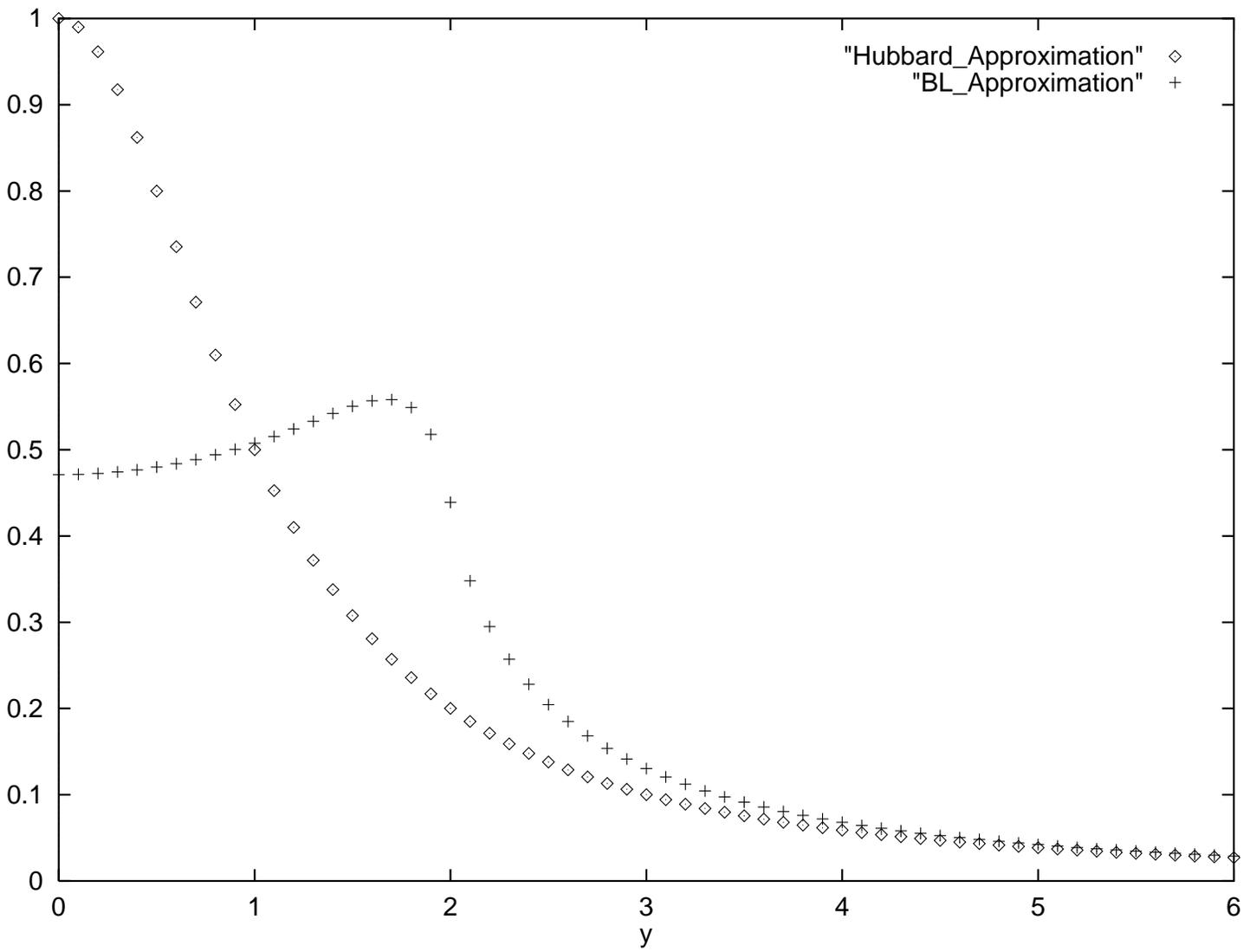}}
\caption{Comparison of the Hubbard approximation and the BL approximation of 
the Coulomb exchange interaction : 
$ \langle \frac{1}{|\vec{k}_1 -\vec{k}_2 + \vec{y}|^2} \rangle$ vs. 
$y$ .}
\label{HBL_graph}
\end{figure}

In this section, we give the contribution of diagrams like those shown in
Fig.~\ref{RingExchange} to the correlation energy at zero temperature. 
Hence
we go beyond RPA in this case. We will show that our method provides excellent
agreement with fully numerical calculations. We also compare our results to
those found in Ref.\cite{bishop} where a coupled cluster formalism has 
been used to get the correlation energy. Our final
results clearly show that our method is more transparent than other previously
used methods. To get to these final results we had to make approximations
along the way. We use two different approximations: the Hubbard approximation
in Ref.\cite{hubbard} and the Bishop-Luhrmann approximation, Ref.\cite{bishop}. We have
found that the former applies well to high values of $r_{s}$ while the latter
applies well to low values of $r_{s}$. From Eq.(118), this amounts to finding
$I_{2}^{exch}$ and the second order exchange term that has already been
calculated by Onsager et al. in Ref.\cite{onsager}. Now we 
show how to calculate
$I_{2}^{exch}$. The calculation is straightforward but special care must be
exercised when it comes to numerically evaluating the final result. First we let%

\begin{equation}
I=\int_{-\infty}^{\infty}dx\int_{0}^{\infty}y^{2}\;dy\ln\left[  1+\frac
{\mathcal{M}(x,y)}{1+\mathcal{H}(x,y)}\right]
\end{equation}
where%

\begin{equation}
\mathcal{M}(x,y)= \frac{16\alpha^{2}r_{s}^{2}}{(2\pi)^{4}}\frac{1}{y^{2}}
\int_{\Gamma}\frac{d^{3}k_{1}d^{3}k_{2}}{|\vec{k_{1}}-\vec{k_{2}}+\vec{y}
|^{2}} \frac{1}{ \left ( -ix+ \frac{\left ( (\vec{k_{1}}+\vec{y})^{2}
-k_{1}^{2} \right )}{2} \right ) \left ( ix+\frac{\left ( (\vec{k_{2}} -
\vec{y}  )^{2} - k_{2}^{2} \right ) }{2} \right )} 
\end{equation}
and where the region of integration $\Gamma$ is given by:

\begin{equation}
\Gamma= \left \{ (k_{1},k_{2})k_{1}<1,\: k_{2}<1,|\vec{y}+\vec{k_{1}
}|>1,\: |\vec{y}-\vec{k_{2}}|>1 \right \} \: .
\end{equation}
It is obvious from the above that we are faced with a daunting task of having
to deal with a 9-D integral inside a logarithmic function which itself 
has to be
integrated over normalized energy $x$ and normalized momentum $y$. Most of the
complications are related to the angle integrals which can not be separated.
To be able to make some progress we have to make an approximation, i.e., we
either assume that $|\vec{k}_{i}+\vec{y}|\approx1$ and average over the angle
between them, i.e., the Hubbard approximation or we use a more sophisticated
approximation like the one proposed by Bishop and Luhrmann (BL). Hubbard's
approximation amounts to the following:
\begin{equation}
\frac{1}{|\vec{k_{1}}-\vec{k_{2}}+\vec{y}|^{2}}\approx\frac{1}{y^{2}+1}\text{
}.
\end{equation}
On the other hand, the BL-approximation is more complicated and the reader is
referred to their appendix in Ref.\cite{bishop}  for 
a discussion of their approximation. The two 
approximations are almost equivalent for large values
of momentum $y$, i.e. $y>2$ in units of Fermi momentum. For values of $y$ less
than 2, both approximations
 are quite different ( see Fig.~\ref{HBL_graph} ). The Hubbard 
approximation seems to apply well for moderately large values of $r_{s}$ while
the BL-approximation applies well for low values of $r_{s}$. In fact the 
BL-approximation incorporates the Fermi statistics of the interacting
particles where it matters most, i.e. in high density situations,
 and this is one 
reason why we found it to be a good approximation to the exchange Coulomb
interaction for low $r_{s}$. The Hubbard approximation should be expected to
be a good approximation to particles near the Fermi surface and where statistics
is not an issue. The case of low density should apply well to this case.
Given either approximation we end up with an
expression for $\mathcal{M}(x,p)$ of the form:
\begin{equation}
\mathcal{M}(x,y)=\frac{16\alpha^{2}r_{s}^{2}}{(2\pi)^{4}}\frac{1}{y^{2}
}f(y)\int_{\Gamma}d^{3}k_{1}d^{3}k_{2}\frac{x^{2}+\frac{1}{4}(y^{2}
+2\vec{k_{1}}\cdot\vec{y})(y^{2}-2\vec{k_{2}}\cdot\vec{y})}{\left (
x^{2}+\frac{1}{4}(y^{2}+2\vec{k_{1}}\cdot\vec{y})^{2} \right ) \left (x^{2}+
\frac{1}{4}(y^{2}-2\vec{k_{2}}\cdot\vec{y})^{2} \right )}
\end{equation}
where the function $f(y)$ is our approximation to the exchange term in
$\mathcal{M}(x,y)$ and is assumed known. \ 

The integrations over $\vec{k_{1}}$ and $\vec{k_{2}}$ are easily found if we
choose to write both vectors in cylindrical coordinates with $\vec{y}$ along
the z-axis, i.e.,
\begin{equation}
\vec{k_{i}}=(\rho_{i},z_{i},\theta_{i}),\: \: \: \: \:  i=1,2 \:.
\end{equation}
Hence the integration over $\vec{k_{1}}$ becomes:
\begin{equation}
\int d^{3}k_{1}\;=\;\int_{0}^{2\pi}\;d\theta_{1}\;\int_{\Gamma^{\prime}}%
\rho_{1}\;d\rho_{1}dz_{1\text{ }},
\end{equation}
where
\[
\Gamma^{\prime}=\left \{ y < 2,\: \: \: -\frac{y}{2} < z_{1} < 1-y,\: \: \:
\left ( 1-(z_{1}+y)^{2} \right ) ^\frac{1}{2} < \rho_{1} < (1-{z_{1}}
^{2})^{^\frac{1}{2}} \right \}
\]
\[
\bigcup \left \{ y<2,\: \: \: 1-y < z_{1} < 1,\: \: \: 0 < \rho_{1} < 
(1-{z_{1}}^{2})^{^\frac{1}{2}} \right \}
\]
\begin{equation}
\bigcup \left \{ y > 2,\: \: \: -1 < z_{1} < 1,\: \: 0 < \rho_{1} <
(1-{z_{1}}^{2})^{\frac{1}{2}} \right \},
\end{equation}
with an equivalent expression for the region of integration over $k_{2}$. The
integrations are now easily carried out. The expression that we get for
$\mathcal{M}(x,y)$ is naturally expressed in terms of $\frac{2x}{y}$ instead
of $x$. So in the following $x$ refers to $\frac{2x}{y}$. We only quote the
final expression here:%

\begin{equation}
\mathcal{M}(x,y)=-16\frac{\alpha^{2}{r_{s}}^{2}}{\pi^{2}}\frac{1}{y^{4}%
}f(y)\text{ }\mathcal{L}(x,y)\text{ },
\end{equation}
with
\begin{eqnarray}
\mathcal{L}(x,y)  & =\left(  -\frac{y^{2}}{16} \left ( \frac{\pi y^{2}
}{2\alpha r_{s}} \right )^{2}\mathcal{H}(x,y) \left (\mathcal{H}
(x,y)-2 \right ) \right.  \nonumber\\
& \mbox{}-\frac{(yx)^{2}}{64}\left\{  y+ \left ( \frac{x}{2}+\frac
{2(1-\frac{y^{2}}{4})}{xy} \right )\left ( \arctan \left ( \frac{2-y}{x}
\right ) - \arctan \left ( \frac{2+y}{x} \right ) \right ) \right.
\nonumber\\
& \mbox{}\left.  +\frac{1}{2}\ln \left[  \frac{(x^{2}+(2-y)^{2})}
{(x^{2}+(2+p)^{2})}\right]  \right \}  \nonumber\\
& \times\left\{  -1+\frac{x}{2} \left ( 1-\frac{y^{2}}{4}+\frac{x^{2}}
{4} \right )\left( \arctan \left ( \frac{2-y}{x} \right ) - \arctan \left (
\frac{2+y}{x} \right ) \right ) + \right.  \nonumber\\
& \left.  \left.  \ln \left[  \frac{ \left ( x^{2}+(2-y)^{2} \right )
^{\frac{1}{2}} \left ( x^{2}+(2+y)^{2} \right )^{\frac{1}{2}}}{x^{2}
}\right]  \right\}  \right)  \Theta(2-y)\nonumber\\
& \mbox{}+\left(  -\frac{y^{2}}{64}(\frac{\pi y^{2}}{2\alpha r_{s}}
)^{2}\mathcal{H}(x,y)(\mathcal{H}(x,y)-1)\right.  \nonumber\\
& \mbox{}-\frac{yx^{2}}{128}\left\{  -4-y\ln\left[  \frac{\left (
x^{2}+(2-y)^{2} \right )^{\frac{1}{2}} \left ( x^{2}+(2+y) \right )
^{\frac{1}{2}}}{x^{2}}\right]  \right.  \nonumber\\
& \mbox{}+\left.  \frac{4}{x} \left ( 1-\frac{y^{2}}{4}+\frac{x^{2}}
{4})(\arctan(\frac{2+y}{x})-\arctan(\frac{-2+y}{x}) \right ) \right\}
\nonumber\\
& \times\left\{  \frac{1}{2}\ln\left[  \frac{ \left ( x^{2}+(2-y)^{2}
\right )^{\frac{1}{2}}\left (x^{2}+(2+y)\right )^{\frac{1}{2}}}{x^{2}
}\right]  \right.  \nonumber\\
& \mbox{}+\left.  \left.  \left.  \frac{2}{x}\frac{1-\frac{y^{2}}{4}}
{y}\left (\arctan(\frac{-2+y}{x})-\arctan(\frac{2+y}{x})\right )\right\}
\right\}  \right)  \Theta(y-2) \: .
\end{eqnarray}
Given this expression for $\mathcal{M}(x,y)$, we can easily write the full
expression for the correlation due to exchange effects in ring diagrams. It is
important to note that we allow for more than one exchange to take place in
each ring diagram. Hence our original integral $I$, can be now evaluated
numerically. So within the above approximation, we were able to reduce our
calculations to a two-dimensional integral:
\begin{equation}
I_{2}^{exch}=\frac{3}{4\pi}\frac{1}{(\alpha r_{s})^{2}}\int_{0}^{\infty}%
dx\int_{0}^{\infty}y^{3}\left(  \ln\left[  1+\frac{\mathcal{M}(x,y)}%
{1+\mathcal{H}(x,y)}\right]  -\mathcal{M}(x,y)\right)  \text{ .}%
\end{equation}
The expression for the function $f(p)$ depends on which approximation we
choose to use. In this formalism it is hard to decide on which one since both
give very close answers to the second order exchange diagram. We have found
that the Hubbard approximation in this instance gives 0.041 $Ryd.$ while the BL
approximation gives 0.049 $Ryd$. Clearly the latter is closer to the true value
of 0.0484 $Ryd.$, but we believe that this is not enough to decide against using the
Hubbard approximation. In fact at low $r_{s}$ we expect that the high energy
part of the exchange energy between particles to dominate and this explains
why we found the BL-approximations to apply well in this range. At high
$r_{s}$ this is not true anymore and in this instance we expect the
BL-approximation to overestimate this exchange of energy between the
particles. This is in fact what we have found. This casts doubts that the
BL-approximation is always better than the Hubbard approximation. As we said
above the second order exchange term is not a good test since it is density
independent. This explains why the Hubbard approximation gives better results
for higher values of $r_{s}$. Below we give some numerical answers to compare
with those obtained using quantum Monte Carlo methods (QMC) \cite{ceperly}.
Our results are obtained using both approximations. For the second order
exchange diagram we have used the true value of 0.0484 Ryd.

\begin{figure}[ptb]
\resizebox{\textwidth}{!}
{\includegraphics[0in,0in][8in,8in]{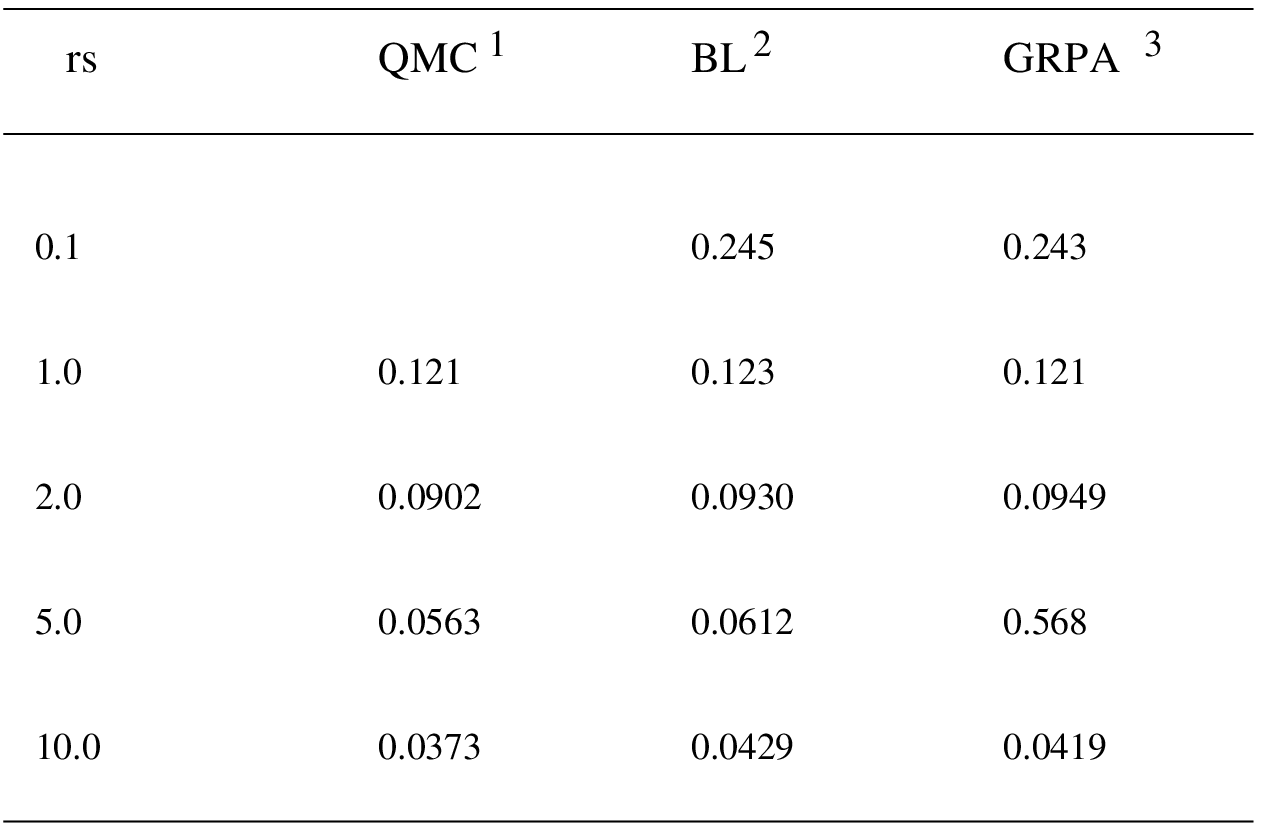}} 
\caption{ Comparison of final results for the correlation energy (in Ryd.) of 
a homogeneous Fermi gas. }
\textsc{\newline 1. Ref.\cite{ceperly} \newline 
2. Ref. \cite{bishop} \newline 3. This paper }
\label{table}
\end{figure}

\noindent Clearly our results, Fig.\ref{table}, agree 
well with previous results. For large
values of $r_{s},$ i.e., $r_{s}$ $\simeq100,$ our results do not compare well
to the QMC results. However this is expected since we need to include
corrections due to the ladder diagrams which are believed to be important at
low densities. This concludes our calculations. \ We have shown the
effectiveness of an effective action approach to electronic systems. In the
next section we shall state the viability of using this formalism in the
nonhonogeneous case and show how it is related to density functional theory.

%

\newpage

\section{Conclusion: The Nonhomogeneous Electron System}

\begin{figure}[ptb]
\resizebox{\textwidth}{!}
{\includegraphics[0in,0in][8in,8in]{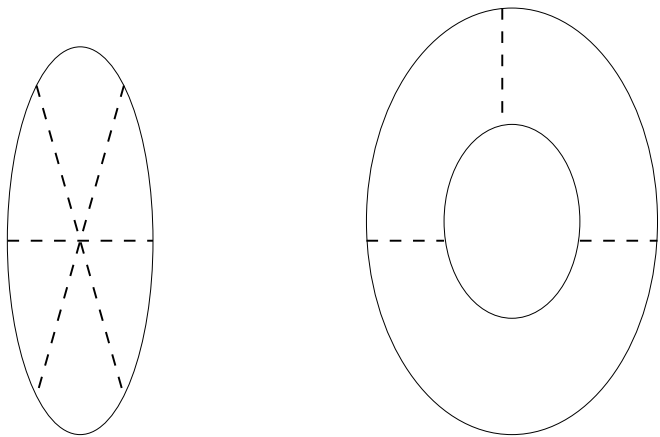}}
\caption{ Diagrams that appear when we include terms of order $\hbar^{6}$ in
$\Gamma$. }
\label{nextTerm}
\end{figure}

In this last section, we would like to go back to the inhomogeneous case and
show how our method gives some of the results known in DFT \cite{kohn}
\cite{sham}. We start by briefly reviewing the basic ideas behind DFT. The
Hamiltonian H is written as the sum of three terms: T, V and U. T is the
kinetic energy term, V is the external potential term and U is the Coulomb
energy term. So we have
\begin{equation}
H=T+V+U  \: \:,
\end{equation}%
\begin{equation}
T=\frac{1}{2}\int\nabla\psi^{\dagger}(r)\nabla\psi(r)d^{3}r\text{ },
\end{equation}%
\begin{equation}
V=\int V(r)\psi^{\dagger}(r)\psi(r)d^{3}r\text{ },
\end{equation}%
\begin{equation}
U=\frac{1}{2}\int\frac{\psi^{\dagger}(r)\psi^{\dagger}(r^{\prime}%
)\psi(r^{\prime})\psi(r)}{|r-r^{\prime}|}dr\;dr^{\prime}\text{ }.
\end{equation}
\newline If $\Psi$ is the ground state, then the density $n(r)$ is given by
\begin{equation}
n(r)=\langle\Psi|\psi^{\dagger}(r)\psi(r)|\Psi\rangle\text{ }.
\end{equation}
The first important fact to note is that $V(r)$ is a unique functional of
$n(r)$, assuming that there is a unique ground state for the system. Now let
the functionals $F[n(r)]$ and $E_{V}[n(r)]$ be defined as follows:%

\begin{equation}
F[n(r)]=\langle\Psi|T+U|\Psi\rangle\text{ },
\end{equation}%
\begin{equation}
E_{V}[n(r)]=\int V(r)n(r)d^{3}r+\;F[n(r)]\text{ }.
\end{equation}
Then, if we assume that there is a one-to-one correspondence between $n(r)$
and $V(r)$, it can be shown that
\begin{equation}
\frac{\delta E_{V}[n]}{\delta n(r)}|_{g.s}=0\text{ },
\end{equation}
with
\[
\int n(r)\;d^{3}r=N\: \: .
\]
\newline This result is called the Hohenberg-Kohn theorem. Next, the Coulomb
energy is isolated from the functional $F[n]$ by introducing a new functional
$G\left[  n\right]  $:
\begin{equation}
F[n]=\frac{1}{2}\int \: d\vec{r}\: d{\vec{r}}^{\prime}
\frac{n(r)\;n(r^{\prime})}{|r-r^{\prime}|}\: + \: G[n]\: ,
\end{equation}%
\begin{equation}
G[n]=\frac{1}{2}\int \:d\vec{r}\: \: \nabla_{r}\nabla_{r^{\prime}}
\: n_{1}(r,r^{\prime
})|_{r=r^{\prime}}+\frac{1}{2}\int\frac{C_{2}(r,r^{\prime})}{|r-r^{\prime}%
|}dr\;dr^{\prime},
\end{equation}
and
\begin{equation}
C_{2}(r,r^{\prime})=n_{2}(r,r^{\prime};r,r^{\prime})-n(r)n(r^{\prime})\text{
}.
\end{equation}
\newline Here $n_{1}(r,r^{\prime})$ is the one-particle density matrix and
$C_{2}(r,r^{\prime})$ is a correlation function defined in terms of the one-
and two-particle density matrices. DFT calculations are essentially centered
around finding good approximations to the functional $G[n]$. This is usually
done by postulating that there is a virtual system of free electrons in an 
external potential with exactly the same density as the interacting
system. The energy is found by finding the eigenfunctions that correspond
to this external potential self consistently. \newline  The
method that was presented here is in fact similar in many ways to the above
ideas expressed in DFT. However there are major differences. Our method 
clearly incorporates the
Hohenberg-Kohn theorem. From Eq.(35), we have the following
\begin{eqnarray}
\frac{\delta E_{gs} \left [ \rho \right ]}{\delta \rho (x,x)} = - Q(x,x) \: . 
\end{eqnarray}
From Eq.(49), the term
\begin{eqnarray}
\int d^{3}x \: V(x) \rho(x,x) \: ,
\end{eqnarray}
can be separated from the energy $E_{gs}$. Hence  if we set $Q(x,x) = 0$ , we
get right away the Hohenberg-Kohn result. In fact we now have  an explicit
expression for $E_{V}[n]$ within perturbation theory. This result also 
shows the one-to-one correspondence between density and external potential
as long as there is a unique solution around $(J,B,Q)=(0,0,0)$ for the 
defining equations, Eqs.(33-35). We have solved these equations only
approximately. We have made two important approximations. The first 
corresponds to the number of diagrams included in $\Gamma$ from the outset.
However from the calculations of the homogeneous case, we see that it 
will probably be the case that including only two diagrams in the inhomogeneous
case will give good results. Including higher order corrections by taking
into account diagrams like the ones in Fig.~\ref{nextTerm} is nevertheless
straightforward if more accurate results are desired. The second approximation
we have made is that in solving for $\rho(x,y)$. We had to linearize 
the equations of motions to be able to solve for $\rho(x,y)$ iteratively. Also
from Eq.(50), we get the
following equivalent result:
\begin{equation}
\nabla^{2}\varphi_{c}(r)=4\pi n(r)\text{ }.
\end{equation}
\newline Clearly our method is a generalization of the Thomas-Fermi 
method  where
$n(r)$ in the last equation is the \emph{true} density of the system. A very
important difference from DFT is that
within this  method we have a systematic scheme for calculating the functional
$G[n]$, Eq.(49). From the above analysis in the homogeneous case, it is 
obvious that the most important contribution to correlation at zero 
temperature comes from calculating the following term:
\begin{eqnarray}
Tr \ln \mathcal{X} = \ln \: det \frac{A^{-1} + e^{2} \rho \rho \mathcal{X} 
+ e^{4} \mathcal{X} \mathcal{X} \rho \rho A \rho \rho}{A^{-1}} \: 
\end{eqnarray}
where
\begin{eqnarray}
\mathcal{X}  \: \equiv \: A^{-1} \: C \: .
\end{eqnarray}
To calculate the determinant of the above operators, we need a basis of
wavefunctions. In practical computations, it is more advantageous to use 
a finite basis set. If we choose as our basis set the wavefunctions
$\left \{ \phi_{i} \right \}_{i=1}^{N}$ with eigenvalues $\epsilon_{i}$
such that
\begin{eqnarray}
\left (-\frac{1}{2} \nabla^{2} + V(x) \right ) \phi_{i}(x) = \epsilon_{i} \phi_{i}(x),
\end{eqnarray}
then the wavefunctions, $\psi_{j}$, of the full interacting system 
of $N$ electrons can 
be written as a linear combination of the above wavefunctions:
\begin{eqnarray}
\psi_{j}(x,t) \: = \: \sum_{i=1}^{N} a_{ji} \phi_{i}(x) e^{-\epsilon_{i} t} \: .
\end{eqnarray}
The coefficients $a_{ji}$ are found self consistently by solving Eq.(56). To 
solve for $\mathcal{X}(x,y)$, we solve Eq.(53) self-consistently. Clearly 
our method is computationally more intensive than DFT. However here we
are on  more secure grounds and our approximations are very well controlled
to any desired accuracy. Plans to apply this method to simple atoms 
are underway. We hope to report on these developments elsewhere. Finally, we 
would like to point out that this method applies equally well to excited
states other than the ground state. The Hartree field of the excited state
is found again by using Eq.(33). If $\varphi_{0}(x)$ is the ground state
solution then $\varphi_{1}(x) = \varphi_{c}(x) + \Delta(x)$ is the 
excited state solution iff
\begin{eqnarray}
\frac{\delta \Gamma}{ \delta \varphi(x)} {\left | \right .}_{\varphi 
= \varphi_{1}} = 0.
\end{eqnarray}
This implies that
\begin{eqnarray}
\int d^{4}y \: \frac{\delta^{2} \Gamma}
{\delta \varphi(x) \delta \varphi(y)} {\left | \right .}_{
\varphi_{0}} \Delta(y) = 0.
\end{eqnarray}
In the above $\Delta(x)$ is assumed small compared to $\varphi_{0}(x)$. This 
latter equation enables us to solve for $\Delta(x)$ and hence the corresponding
$\rho(x,y)$. Clearly effective actions methods are in general  superior to 
conventional methods as is well known in field theory. We expect this 
method to apply equally well to atomic and molecular systems as we have shown
to be the case for homogeneous systems.

\end{document}